\documentclass[twocolumn]{svjour3} 

\smartqed 
\PassOptionsToPackage{table}{xcolor}
\usepackage{tikz}
\usetikzlibrary{shapes}
\usepackage{graphicx}
\usepackage{subfigure}
\usepackage{lmodern}
\usepackage{booktabs}
\usepackage{algorithm,algpseudocode}
\usepackage{float}
\usepackage{amssymb}
\usepackage{amsmath}
\usepackage{bm}
\usepackage{hyperref}
\usepackage{natbib}
\usepackage{lineno}
\usepackage{soul}
\usepackage{placeins}

\usepackage{tabularx}
\usepackage[range-phrase = --,range-units = single, product-units = power, per-mode = symbol]{siunitx}
\DeclareSIUnit\pixel{px}
\DeclareSIUnit\voxel{vox}

\usepackage{multirow}

\usepackage{comment}

\newlength{\tabind}
\newcommand*{\T}{\mathsf{T}}

\allowdisplaybreaks

\begin{document}
\definecolor{color1}{rgb}{0.267004, 0.004874, 0.329415}
\definecolor{color2}{rgb}{0.229739, 0.322361, 0.545706}
\definecolor{color3}{rgb}{0.127568, 0.566949, 0.550556}
\definecolor{color4}{rgb}{0.369214, 0.788888, 0.382914}
\definecolor{color5}{rgb}{0.993248, 0.906157, 0.143936}

\newcommand{\markerone}{\raisebox{0.0pt}{\tikz{\node[draw,scale=0.6,circle,fill=color1, line width=0.02cm](){};}}}
\newcommand{\markertwo}{\raisebox{0pt}{\tikz{\node[draw,scale=0.6,regular polygon, regular polygon sides=4,fill=color2, line width=0.02cm](){};}}}
\newcommand{\markerthree}{\raisebox{-1.0pt}{\tikz{\node[draw,scale=0.6,regular polygon, regular polygon sides=4,fill=color3,rotate=45, line width=0.02cm](){};}}}
\newcommand{\markerfour}{\raisebox{-0.5pt}{\tikz{\node[draw,scale=0.45,regular polygon, regular polygon sides=3,fill=color4,rotate=180, line width=0.02cm](){};}}}
\newcommand{\markerfive}{\raisebox{-0.7pt}{\tikz{\node[draw,scale=0.35,star,star point ratio=2.5,fill=color5, line width=0.02cm](){};}}}

\sloppy
\title{A meshless data-tailored approach to compute statistics from scattered data with adaptive radial basis functions.}

\author{
    Damien Rigutto \and 
    Manuel Ratz \and
    Miguel A. Mendez
}

\institute{
    Damien Rigutto \at
    Environmental and Applied Fluid Dynamics, von Karman Institute for Fluid Dynamics, Sint-Genesius-Rode, Belgium \\
    Transfers, Interfaces \& Processes (TIPs) Laboratory, École Polytechnique de Bruxelles, Université Libre de Bruxelles, Brussels, Belgium \\
    \email{damien.rigutto@vki.ac.be}
    \and
    Manuel Ratz \at
    Environmental and Applied Fluid Dynamics, von Karman Institute for Fluid Dynamics, Sint-Genesius-Rode, Belgium \\
    Aero-Thermo-Mechanics Laboratory, École Polytechnique de Bruxelles, Université Libre de Bruxelles, Brussels, Belgium \\
    \email{manuel.ratz@vki.ac.be}
    \and
    Miguel A. Mendez \at
    Environmental and Applied Fluid Dynamics, von Karman Institute for Fluid Dynamics, Sint-Genesius-Rode, Belgium \\
    Aero-Thermo-Mechanics Laboratory, École Polytechnique de Bruxelles, Université Libre de Bruxelles, Brussels, Belgium \\
    Experimental Aerodynamics and Propulsion Lab, Universidad Carlos III de Madrid, Leganés, Spain \\
    \email{miguel.alfonso.mendez@vki.ac.be}
}

\date{\today}

\maketitle

\abstract{
Constrained radial basis function (RBF) regression has recently emerged as a powerful meshless tool for reconstructing continuous velocity fields from scattered flow measurements, particularly in image-based velocimetry. However, existing formulations based on isotropic kernels often suffer from spurious oscillations in regions with sharp gradients or strong flow anisotropy. 
This work introduces an anisotropic, gradient-informed, and adaptively sampled extension of the constrained RBF framework for regression of scattered data. Gradient information is estimated via local polynomial regression at collocation points, smoothed, and used to (1) re-sample data, maximizing sampling density near steep gradients while downsampling in smooth regions, and (2) construct a local anisotropic metric that shapes each basis function according to the flow directionality. In addition, a gradient-informed regularization is introduced by embedding observed gradients into the least-squares system as weighted soft constraints. 
The resulting formulation is fully meshless, linear, and computationally efficient, while significantly improving reconstruction quality in challenging regions. The method is evaluated on both synthetic and experimental datasets, including direct numerical simulation (DNS) data of a turbulent channel and time-resolved particle tracking velocimetry of a turbulent jet. Results show that the proposed approach outperforms isotropic and gradient-free RBF formulations in accuracy, smoothness, and physical consistency---particularly near shear layers and boundaries---while reducing the number of bases by an order of magnitude. To support the application, we have created a repository (\url{https://github.com/mendezVKI/SPICY_VKI}) that provides access to the investigated datasets.
}

\vspace{-5mm}

\section{Introduction}
\label{Sec:Intro}

Constrained radial basis function (RBF) regression methods 
have recently emerged as tools for data assimilation from scattered 
measurements, such as those obtained through Lagrangian particle tracking 
velocimetry (LPT) \citep{Schroeder2023}. Established approaches such as VIC+ 
or VIC\# \citep{Schneiders2016,Scarano2022,Jeon2022}, as well as the 
spline-based FlowFit methods \citep{Gesemann2016,Godbersen2024}, were primarily designed to map scattered data onto a structured grid, either explicitly (VIC-type methods) or through grid-defined basis functions (FlowFit). 
In contrast, constrained RBF methods seek to reconstruct an analytic 
representation of the underlying velocity field directly from the scattered data, and to carry out all subsequent post-processing — differentiation (including pressure reconstruction \citep{Ratz2022,Sperotto2022a,Sperotto2024a,Li2025}), modal decomposition \citep{Tirelli2025a,Ratz2025}, or turbulence statistics \citep{Ratz2024}—to be performed without ever relying on a grid. \textcolor{black}{This allows for handling complex geometries and arbitrary domains, analytic differentiation, the absence of grid-induced anisotropy, and the ability to perform super-resolution or locally refined evaluations without any remeshing.}

A meshless formulation is also offered by recent neural-network-based approaches, such as sinusoidal representation networks (SIRENs; \citealp{Miotto2025}) and physics-informed neural networks (PINNs; \citealp{Zhou2023b,cai2024}), which have achieved high accuracy in recent data-assimilation challenges \citep{Zhou2024}. These approaches construct continuous, differentiable representations of the flow field and can embed physical laws as soft constraints during training. However, methods based on neural networks require solving high-dimensional nonlinear optimization problems for the training, involving extensive hyperparameter tuning, and regularization to balance physical and regression losses. In contrast, constrained RBF regression is a linear method that solves a single linear system while allowing physical constraints to be imposed as hard conditions. This makes RBF-based approaches substantially more efficient and robust, and thus well suited for experimental fluid mechanics. Moreover, the ensemble formulation of \citet{Ratz2024} provides a natural framework for statistical flow analysis, enabling ensemble averages and higher-order statistics to be obtained within a single regression.

Nevertheless, despite these advantages, two major limitations remain. The first concerns the difficulty of accurately representing sharp gradients and highly confined flow structures, such as those occurring in shear or boundary layers. These regions often induce spurious oscillations reminiscent of the Runge phenomenon in polynomial interpolation \citep{Fornberg2007, Boyd2010}. The second limitation arises in the computation of statistics from the ensemble formulation itself: when combining many snapshots into a single regression problem, the size of the resulting system can become computationally prohibitive. This work addresses both issues through a combination of (1) adaptive sub-sampling and collocation, (2) anisotropic stretching of the bases, and (3) gradient penalization.

Adaptive refinement has been widely explored in RBF methods, both for 
interpolation and for solving partial differential equations (PDEs). In 
interpolation, \citet{Driscoll2007} introduced the adaptive residual subsampling method, where nodes are iteratively refined or pruned based on local residuals, enabling accurate reconstruction of localized features with a minimal number of  basis functions. For PDE problems, \citet{Sarra2005} demonstrated that the grid-free nature of RBFs allows adaptive relocation of collocation points to track steep gradients in time-dependent systems, while \citet{Cavoretto2020} proposed a two-stage adaptive scheme combining cross-validation and residual-based refinement to efficiently handle elliptic problems with sharp spatial variations. 

The adaptive sub-sampling strategy proposed in this work, however, follows a 
different philosophy. Rather than generating new points, it prunes an existing 
dense ensemble of measurements through an adaptive sub-sampling guided by the 
local gradient magnitude of the inferred field. This gradient-based criterion 
defines a probability density for sample selection, retaining more points in 
regions of high spatial variability and fewer in smooth areas. Collocation 
points are then distributed using a variable-density Poisson disk sampling 
strategy \citep[e.g.][]{Slak2019, Dwork2021, Lawrence2024}, ensuring that the 
RBF bases remain well separated and uniformly cover the domain without excessive 
overlap.
The proposed anisotropic RBF formulation draws inspiration from techniques 
developed in PDE solvers \citep{King2020, Cheng2018} and partition of unity 
methods (PUM), where subdomain shapes are adapted to directional features 
\citep{Cavoretto2019}. Similar concepts have been applied to image interpolation 
to preserve sharp transitions \citep{Casciola2006, casciola2010} and to the 
modeling of free-form surfaces \citep{Maksimovic2016}. \citet{Beatson2010} 
provided a rigorous analysis of the benefits of aligning anisotropic bases with local directionality, demonstrating significant accuracy gains in interpolation problems. In the context of statistical analysis of scattered velocimetry data, a related idea is the anisotropic binning of \citet{Raiola2020}, which elongates local averaging regions following the least-squares polynomial regressions of \citet{Agueera2016}. Building on these ideas, the present work incorporates anisotropic stretching within the constrained RBF regression framework.

Finally, the proposed gradient penalization acts as a regularization term that controls the smoothness of the reconstructed field. This approach is inspired by gradient-enhanced RBF interpolation \citep{Laurent2019} and by Gaussian process regression formulations \citep{Bhaduri2020, Maeaettae2021}, where derivative information improves local adaptivity and reduces uncertainty near steep transitions. Here, the gradient penalty stabilizes the regression in regions of strong gradients while remaining consistent with the ensemble-based statistical formulation.

The remainder of this article is structured as follows. 
Section~\ref{sec.methodology} reviews the constrained RBF framework and its extensions. 
Section~\ref{sec.test} describes the test cases used for benchmarking. 
Section~\ref{sec.results} presents and discusses the results, and 
Section~\ref{sec.conclusions} concludes the paper with a summary and outlook.

\section{Methodology}\label{sec.methodology}

We start with a brief review of the fundamentals of constrained and penalized RBF regression in Sec.~\ref{sec2p1}. Section~\ref{sec2p2} introduces the methodology for anisotropic adaptation, while Sec.~\ref{sec2p3} presents the gradient-informed extension. For clarity and simplicity, the method is described in two dimensions to streamline the notation; however, its extension to three dimensions is straightforward. \textcolor{black}{We consider datasets of scattered measurements, as typically encountered in particle-based velocimetry, assuming a sampling density sufficient for standard reconstruction of flow gradients. Our goal is not to probe the limits of RBF regression under extreme sparsity, but to enhance its accuracy and efficiency through adaptive, anisotropic formulations, given a baseline regime where the reconstruction problem is well-posed.}

\subsection{The constrained/penalized RBF framework}\label{sec2p1}

We consider the reconstruction of a scalar field \( f: \mathbb{R}^2 \to \mathbb{R} \) from a set of scattered observations \(\{(\mathbf{x}_{\ast,i}, f_{\ast,i})\}_{i=1}^{n_P}\), with \(\mathbf{x}_{\ast,i} \in \mathbb{R}^2\) and \(f_{\ast,i} = f(\mathbf{x}_{\ast,i})\). The function is approximated using a linear combination of $n_B$ radial basis functions:
\begin{equation}
    f(\bm{x}) \approx \tilde{f}(\bm{x}) = \sum_{m=1}^{n_B} w_m \, \gamma_m(\bm{x};\, \mathbf{x}_{c,m}, c_m)\,,
    \label{eq:rbf_model}
\end{equation}
where \(\mathbf{x}_{c,m}\) and \(c_m\) are the collocation point and shape parameter of the $m$-th RBF $\gamma_m$ (fully defined in Section~\ref{sec2p2}), and \(\bm{x} = (x, y)\) is an arbitrary point in space. The semicolon distinguishes the variables from the parameters. The vector of weights \(\mathbf{w}=[w_1,w_2,\dots,w_{n_B}]^\T \in \mathbb{R}^{n_B}\) is determined from the available data.

Following \citet{Sperotto2022a} and \citet{Ratz2024}, we approximate each component of a vector field independently using separate scalar regressions. Therefore, enforcing vector-valued constraints (e.g., divergence-free conditions) typically requires coupling the velocity components into a larger system. However, since this work focuses exclusively on scalar constraints (e.g., no-slip or non-penetration), each component can be treated independently.  
Alternative formulations include representing the vector field using scalar potentials (e.g., stream or potential functions), or employing vector-valued basis functions that intrinsically enforce structural properties such as freedom of divergence \citep{Yang2014,Fuselier2015,Li2025}.

Defining \(\mathbf{\Gamma}(\mathbf{X}_\ast) \in \mathbb{R}^{n_P \times n_B}\) as the matrix collecting the evaluations of the basis functions at the sample locations \(\mathbf{X}_\ast = [\mathbf{x}_{\ast,1}, \dots, \mathbf{x}_{\ast,n_P} ] \in \mathbb{R}^{2\times n_P}\), and \(\mathbf{f}_\ast \in \mathbb{R}^{n_P}\) as the vector of observed values, \eqref{eq:rbf_model} can be written in matrix form as:
\begin{equation}
    \mathbf{f}(\mathbf{X}_\ast) \approx \tilde{\mathbf{f}}(\mathbf{X}_\ast) = \mathbf{\Gamma}(\mathbf{X}_\ast) \mathbf{w}\,,
    \label{eq:rbf_matrix}
\end{equation}
and the weight vector \(\mathbf{w}\) is obtained by solving the least-squares problem:
\begin{equation}
    \min_{\mathbf{w}} \left\| \mathbf{\Gamma}(\mathbf{X}_\ast) \mathbf{w} - \mathbf{f}_\ast \right\|_2^2\,,
    \label{eq:least_squares}
\end{equation} where $|| \cdot ||_2$ is the $l_2$ norm of a vector.

Because the regression model \eqref{eq:rbf_matrix} is linear in \(\mathbf{w}\), both hard and soft constraints can be easily incorporated when these are also linear. Linear constraints can be used to enforce boundary conditions (e.g., no-slip or symmetry) or linear differential conditions (e.g., solenoidal constraints for incompressible flows). Hard constraints are introduced by augmenting the least-squares problem with equality conditions:
\begin{equation}
    \begin{aligned}
        & \min_{\mathbf{w}} \left\| \mathbf{\Gamma}(\mathbf{X}_\ast) \mathbf{w} - \mathbf{f}_\ast \right\|_2^2 \\
        & \text{s.t.} \quad \mathbf{C}_1(\mathbf{X}_1) \mathbf{w} = \mathbf{c}_1, \quad \mathbf{C}_2(\mathbf{X}_2) \mathbf{w} = \mathbf{c}_2, \quad \dots
    \end{aligned}
    \label{eq:hard_constraints}
\end{equation}
where \(\mathbf{C}_1\) and \(\mathbf{C}_2\) are constraint operators (e.g., function values or derivatives evaluated at specific points \(\mathbf{X}_1,\,\mathbf{X}_2\)), and \(\mathbf{c}_1,\,\mathbf{c}_2\) are the corresponding target values.
This leads to a quadratic programming problem that can be solved via the Karush-Kuhn-Tucker (KKT) conditions, yielding a saddle-point system of the form \citep{Sperotto2022a}:
\begin{equation}
\begin{pmatrix}
\mathbf{\Gamma}^\T \mathbf{\Gamma} & \mathbf{C}^\T \\
\mathbf{C} & \mathbf{0}
\end{pmatrix}
\begin{pmatrix}
\mathbf{w} \\
\boldsymbol{\lambda}
\end{pmatrix}
=
\begin{pmatrix}
\mathbf{\Gamma}^\T \mathbf{f}_\ast \\
\mathbf{c}
\end{pmatrix},
\label{eq:kkt_system}
\end{equation}
where \(\mathbf{C}=[\, \mathbf{C}_1 \;\; \mathbf{C}_2\;\; \dots \, ]\) is the concatenated block matrix of constraints, and \(\boldsymbol{\lambda}\) is the vector of associated Lagrange multipliers.  
Hard constraints increase the size of the problem, and hence the computational cost, since they introduce additional unknowns. When exact enforcement is not required, {soft constraints} can instead be introduced by augmenting the objective function with penalty terms, leading to a {penalized regression} formulation:
\begin{equation}
    \min_{\mathbf{w}} 
    \left\| \mathbf{\Gamma}(\mathbf{X}_\ast) \mathbf{w} - \mathbf{f}_\ast \right\|_2^2
    + \sum_{i} \alpha_i \left\| \mathbf{C}_i (\mathbf{X}_i) \mathbf{w} - \mathbf{c}_i \right\|_2^2\,,
    \label{eq:soft_constraints}
\end{equation}
where \(\alpha_i > 0\) are user-defined regularization parameters, and the operators \(\mathbf{C}_i\) define the soft constraint conditions.  
In contrast to the hard-constrained formulation, penalty terms add equations without introducing new unknowns. To balance accuracy and computational efficiency, \citet{Sperotto2022a} and \citet{Ratz2024} employ a hybrid approach combining hard and soft constraints. In the present work, we adopt a purely penalization-based (soft-constraint) formulation for boundary conditions, while noting that extension to hard constraints is straightforward.

\begin{algorithm}
\caption{Adaptive Gradient-Based Basis Construction}
\begin{algorithmic}[1]
    \State \textbf{Inputs:}
    \Statex Full data $\{(\bm{x}_{\text{full},i}, f_{\text{full},i})\}_{i=1}^{n_P,\mathrm{full}}$, regression number of points $n_P$, neighbour count $n_F$, gradient subsample size $n_G$, gradient spread $\sigma_S$, minimal sampling probability $\text{min}(p(\mathbf{x}_{*,i}))$, Poisson disk distances $r_{\min}, r_{\max}$, basis parameter $\varepsilon_R$, maximum aspect ratio $\text{AR}_{\max}$\,.
    \vspace{0.5em}
    \State \textbf{(1) Gradient Estimation}
    \State Randomly sample $\mathbf{X_\text{full}}=[\bm{x}_{\text{full},1}\dots]\in\mathbb{R}^{2\times n_{P,\text{full}}}$ to select $\mathbf{X}_*\in\mathbb{R}^{2\times n_P}$.
    \State Randomly re-sample $\mathbf{X_*}=[\bm{x}_{*,1}\dots]\in\mathbb{R}^{2\times n_P}$ to select $\mathbf{X}_g\in\mathbb{R}^{2\times n_G}$, with $n_G\ll n_P$.
    \For{each $\bm{x}_{g,k}\in\mathbf{X}_g$}
        \State Find neighbourhood $\mathcal{N}_k$: the $n_F$ nearest neighbours of $\bm{x}_{g,k}$
        \State Compute weights $\omega_j=\exp(-\|\bm{x}_j-\bm{x}_{g,k}\|_2^2/\sigma_F^2)$
        \State Solve weighted least-squares polynomial fit
        \State Compute gradient $\mathbf{g}_{g,k}$ from fitted coefficients
    \EndFor
    \State Create matrix $\mathbf{G}_g=\{\mathbf{g}_{g,k}\}\in\mathbb{R}^{2\times n_G}$.

    \vspace{0.5em}
    \State \textbf{(2) Adaptive downsampling}
    \State Smooth $|\mathbf{G}_g|$ with a Gaussian of standard deviation $\sigma_S$, normalize the result to $[0,1]$. Recover the sign by smoothing $\mathbf{G}_g$ with the same Gaussian and taking its sign, then combine the normalized magnitude with the recovered sign to obtain $\tilde{\mathbf{G}}_g \in [-1, 1]$.
    \State Interpolate $\tilde{\mathbf{G}}_{g}^\mathrm{interp}(\bm{x})$ using nearest neighbour interpolation
    \State Construct sampling probability estimator from \eqref{eq:final_pdf} 
    \State Clip to ensure minimal probability 
    \State Resample $\mathbf{X_*}$ using $p(\mathbf{x}_{*,i})$
    \State Recompute $\mathbf{G}_g$, $\tilde{\mathbf{G}}_g$ and $\tilde{\mathbf{G}}_g^{\text{interp}}$ with adaptive data

    \vspace{0.5em}
    \State \textbf{(3) Adaptive Basis Collocation}
    \State Compute spacing function \\ 
    \hspace{5mm} $r_c(\bm{x})=(r_{\max}-r_{\min})(1-\sqrt{\|\tilde{\mathbf{G}}_{g}^\mathrm{interp}(\bm{x})\|_2})+r_{\min}$
    \State Run variable-density Poisson disk sampling $\to \mathbf{X}_c$
    \For{each $\bm{x}_{c,m}\in\mathbf{X}_c$}
        \State Set basis scale $c_m$ such that $\gamma_m=\varepsilon_R$ at half nearest-neighbor distance
        \State Initialize isotropic metric $\mathbf{M}_m^{iso}=c_m^2\mathbf{I}$
    \EndFor

    \vspace{0.5em}
    \State \textbf{(4) Anisotropic Adaptation}
    \For{each $\bm{x}_{c,m}\in\mathbf{X}_c$}
        \State Evaluate $\tilde{\mathbf{g}}_m=\tilde{\mathbf{G}}_{g}^\mathrm{interp}(\bm{x}_{c,m})$
        \State Compute aspect ratio $\text{AR}_m = \|\tilde{\mathbf{g}}_m\|_2(\text{AR}_{\max}-1)+1$
        \State Directions: $\mathbf{t}_m=\tilde{\mathbf{g}}_m/\|\tilde{\mathbf{g}}_m\|_2$, $\mathbf{n}_m=[-t_{m,y},\,t_{m,x}]^\top$
        \State Raw metric: $\mathbf{M}_m^\text{raw}=\text{AR}_m\,\mathbf{t}_m\mathbf{t}_m^\top+\frac{1}{\text{AR}_m}\mathbf{n}_m\mathbf{n}_m^\top$
        \State Scale factor: $s_m=(\det(\mathbf{M}_m^\text{iso})/\det(\mathbf{M}_m^\text{raw}))^{1/2}$
        \State Final metric: $\mathbf{M}_m=s_m\,\mathbf{M}_m^\text{raw}$
    \EndFor
\end{algorithmic}
\label{Algo_1}
\end{algorithm}

\subsection{Anisotropic RBFs}\label{sec2p2}

In the most common isotropic RBF formulation, the $m$-th basis function \eqref{eq:rbf_model} centred at $\mathbf{x}_{c,m}$ is defined in terms of the Euclidean distance to the evaluation point $\bm{x}\in\mathbb{R}^2$:
\begin{equation}
 \gamma_m(\bm{x} ;\mathbf{x}_{c,m},\mathbf{h}_m) = \gamma\left( \| \bm{x} - \mathbf{x}_{c,m} \|^2_2 ;\,\mathbf{h}_m \right)\,,
\end{equation} where $\gamma: \mathbb{R}^2 \to \mathbb{R}$ is a scalar radial kernel, such as a Gaussian, inverse multiquadric, or thin-plate spline, and $\mathbf{h}_m$ is a vector of hyper-parameters controlling the shape of the $m$-th basis. This isotropic formulation implies that the support of each basis function is spherically symmetric. For instance, the most common isotropic Gaussian basis reads: 
\begin{equation}
\label{gaussian_rbf}
\gamma_m(\bm{x};\mathbf{x}_{c,m},c_m) = \exp\left(-c_m^2 \| \bm{x} - \mathbf{x}_{c,m} \|^2_2\right)\,,
\end{equation} where $c_m$ is the only hyper-parameter besides the collocation point $\mathbf{x}_{c,m}$.

To better capture directional features such as sharp gradients, the anisotropic generalization is obtained by introducing a locally defined anisotropic metric $\mathbf{M}_m$
\begin{equation}
    \gamma_m(\bm{x};\mathbf{x}_{c,m},\mathbf{M}_m ) = \exp\left( - \|\mathbf{M}_m (\bm{x} - \mathbf{x}_{c,m}) \|_2^2  \right)\,,
    \label{eq:anisotropic_rbf_general}
\end{equation}
where $\mathbf{M}_m \in \mathbb{R}^{2 \times 2}$ is a symmetric positive-definite matrix that encodes a local anisotropic stretching at the center $\mathbf{x}_{c,m}$. The anisotropic adaptation is built starting from an initially isotropic basis, hence  this metric is defined as a function of the initial $c_m$, that primarily controls  the size of the basis. The matrix $\mathbf{M}_m$ thus defines a local deformation of space such that distances are contracted or expanded along preferred directions, reshaping the basis function to align with local features of the target field. This metric is defined such that setting \( \mathbf{M}_m = c_m^2 \mathbf{I} \) gives back the isotropic case. In this work, the matrix is constructed from a local estimate of the target function gradient field, whose computation is described below.

The complete procedure for placing, sampling, and anisotropically deforming the basis functions is summarized in Algorithm~\ref{Algo_1}. This essentially prepares the basis and constructs the matrix \(\mathbf{\Gamma}(\mathbf{X}_\ast)\) in \eqref{eq:rbf_matrix}. The process consists of four main steps: (1) gradient estimation from a subset of the dataset, (2) adaptive re-sampling, (3) adaptive basis collocation, and (4) anisotropic basis deformation. The following subsections provide detailed explanations of each of these steps.

\vspace{1mm}\textbf{Step 1: Gradient estimation (lines 3 to 11).} An estimate of the data gradient is required multiple times in the methodology. For this purpose, a reduced set of points $\mathbf{X}_{g} \in \mathbb{R}^{2 \times n_G}$ is randomly subsampled from the full set $\mathbf{X}_{*}$. The estimation is performed using a local polynomial regression. In 2D, the general second-order polynomial fit reads
\begin{equation}
f(\bm{x})\approx  \boldsymbol{p}(\bm{x})^\T \mathbf{a},
\label{eq:poly_vector_form}
\end{equation}
with 
\begin{equation}
\boldsymbol{p}(\boldsymbol{x})^\T = (1\,\ x\,\ y\,\ x^2\,\ xy\,\ y^2)\,,
\label{eq:poly_basis_vector}
\end{equation}
and $\{a_i\}_{i=1}^6$ the set of coefficients stored in the vector $\mathbf{a}$. Accordingly, the gradient vector is approximated as $\nabla f(x, y) \approx [\, a_2 + 2 a_4 x + a_5 y,\;\; a_3 + a_5 x + 2 a_6 y \,]^\T$. The unknowns $\mathbf{a}$ are obtained by solving a local least-squares problem that minimizes the weighted discrepancy between the polynomial model and the known function values at neighbouring points:
\begin{equation}
\min_{\mathbf{a}} \sum_{j \in \mathcal{N}_m} \left( f_j - \boldsymbol{p}(\mathbf{x}_j)^\T \mathbf{a} \right)^2 
\exp\left( -\frac{\|\mathbf{x}_j - \mathbf{x}_{g,k}\|^2_2}{\sigma_F^2} \right),
\label{eq:least_squares_polyfit}
\end{equation}
where $\mathcal{N}_m \in \mathbb{R}^{n_F}$ denotes the neighbourhood of $\mathbf{x}_{g,k}$, and the exponential weight gives more importance to the closest points. The number of points $n_F$ and the weight spreading parameter $\sigma_F$ control the locality of the regression and must balance the numerical stability of the least-squares problem with the resolution required to capture local variations in the target field. The value of $\sigma_F$ is set to obtain a weight of 0.5 at the average distance of the $n_F$ neighbouring points to the centre. The gradient estimation on $\mathbf{X}_g$ is denoted $\mathbf{G}_g$. 

In addition, this step provides a local measure of the sampling density. For each point $\mathbf{x}_{g,k}$ we define a neighbourhood radius
\begin{equation}
r_{g,k} = \max_{j \in \mathcal{N}_m} \|\mathbf{x}_j - \mathbf{x}_{g,k}\|_2,
\end{equation}
i.e.\ the distance to the farthest neighbour in $\mathcal{N}_m$. Under locally quasi-uniform sampling, larger values of $r_{g,k}$ correspond to lower local sampling density. The set $\{r_{g,k}\}$ is used in the following step to bias the selection of points toward sparsely sampled regions.

\vspace{1mm}

\textcolor{black}{
\textbf{Step 2: Adaptive downsampling (lines 13 to 18).} 
This step aims at reducing the size of the regression problem by selecting a subset of points from an already available dataset, while preserving resolution in the regions that matter most. It assumes that the initial dataset is sufficiently informative to resolve the relevant flow gradients, and focuses on identifying an optimal subset for the RBF approximation.}

The downsampling strategy combines two criteria, both derived from the information computed in Step~1: (i) the local gradient magnitude estimate  $\mathbf{G}_g$, which identifies areas of strong spatial variation, and (ii) the local sampling density, inferred from the neighbourhood radius $r_{g,k}$, which highlights sparsely populated regions.

The gradient criterion is based on the idea that regions of high gradients generally need more samples than regions of low gradients. This is required for two reasons. First, because high gradients require small bases, which in turn demand denser sampling to ensure sufficient support. Second, because regions of high gradients are usually associated with strong turbulence and thus require more samples for statistical convergence. To moderate the influence of the gradient on the sampling anisotropy, the absolute gradient values $|\mathbf{G}_g|$ obtained in the previous step are low-pass filtered using a Gaussian kernel with standard deviation $\sigma_S$. 

\textcolor{black}{It is worth stressing that the smoothed gradient does not aim to reproduce local gradient fields but to provide a robust indicator for guiding the adaptive downsampling and basis construction.}

Since the smoothing is applied to the absolute values, gradients of opposite sign do not cancel each other during filtering.
However, this operation removes the sign information associated with the orientation of $\mathbf{G}_g$. To recover it, the sign is recomputed using the same smoothing procedure but applied directly to $\mathbf{G}_g$ instead of $|\mathbf{G}_g|$.
Combining the recovered sign with the smoothed magnitude yields the final quantity $\tilde{\mathbf{G}}_g$, a sign-consistent, smoothed version of $\mathbf{G}_g$ evaluated at the points $\mathbf{X}_g$. A nearest-neighbour interpolant, denoted $\tilde{\mathbf{G}}_{g}^{\mathrm{interp}}$, is constructed from $\tilde{\mathbf{G}}_g$. It is then used to form the gradient-based probability density function $p_G(\mathbf{x}_{*,i})$ by interpolating the smoothed gradient estimates onto the full set $\mathbf{X}_*$. The contribution of the gradients to the sampling pdf is then
\begin{equation}
p_G(\mathbf{x}_{*,i}) 
= \frac{\|\tilde{\mathbf{G}}_{g}^{\mathrm{interp}}(\mathbf{x}_{*,i})\|_2}
       {\sum_{r=1}^{n_P} 
        \|\tilde{\mathbf{G}}_{g}^{\mathrm{interp}}(\mathbf{x}_{*,r})\|_2 }.
\label{eq:pG}
\end{equation}

Setting a large smoothing kernel $\sigma_S$ drives the smoothed gradients $\mathbf{\tilde{G}}_g(\mathbf{X}_g)$ towards a constant and thus recovers the uniform sampling criterion.

Similarly, the density criterion ensures that re-sampling does not amplify existing non-uniformities in the data. If sparse regions were sampled even less, large gaps would appear and local RBF bases could become poorly supported. By using the local radius $r_{g,k}$ as a density indicator, the procedure favours points in sparsely populated areas and down-weights those in already dense regions, thereby maintaining a balanced and representative subset.
Therefore, a density-based probability $p_D(\mathbf{x}_{*,i})$ is defined 
using a nearest neighbour interpolation of the local radius $r_g^\text{interp}(\bm{x})$. We set
\begin{equation}
p_D(\mathbf{x}_{*,i}) 
= \frac{r_g^\text{interp}(\mathbf{x}_i)^2}{\sum_{r=1}^{n_P} r_g^\text{interp}(\mathbf{x}_r)^2}.
\label{eq:pD}
\end{equation}

The final sampling probability is obtained by combining the two criteria:
\begin{equation}
p(\mathbf{x}_{*,i}) \propto p_G(\mathbf{x}_{*,i}) \, p_D(\mathbf{x}_{*,i}),
\qquad
\sum_{i=1}^{n_P} p(\mathbf{x}_{*,i}) = 1.
\label{eq:final_pdf}
\end{equation}
This multiplicative form favours points that simultaneously exhibit strong gradients and low local sampling density.

The whole procedure is enhanced by two additional measures. First, the final distribution is clipped to a prescribed minimum value and subsequently renormalized, so that every point retains a probability of at least $p_{\text{min}}$ (here taken as 0.50) while the distribution remains properly normalized. This helps avoid undersampling regions where the gradient is too small or the density is too large. Second, the identified points are used to repeat both Step 1 and Step 2, yielding usually better gradient estimation, and thus a refined sampling.

\vspace{1mm}

\textbf{Step 3: Adaptive Collocation of Basis (lines 20 to 26).} Following the data sampling, the bases are collocated in an adaptive manner, using the smoothed gradient function $\mathbf{\tilde{G}}_{g}^\mathrm{interp}(\bm{x})$ as an indication of the collocation point density. The collocation is performed using a {variable density Poisson disk sampling} algorithm as described in \citet{Dwork2021}, with a density defined according to $\mathbf{\tilde{G}}_g^\mathrm{interp}$. The main parameter of the algorithm is the minimum inter-point distance $r_c(\bm{x})$, which is mapped from the gradient to a range of minimal and maximal distances, $r_\mathrm{min}$ and $r_\mathrm{max}$, as follows:
\begin{equation}
    r_c(\bm{x}) = (r_\mathrm{max} - r_\mathrm{min}) \biggl(1-\sqrt{\|\mathbf{\tilde{G}}_{g}^\mathrm{interp}(\bm{x})\|_2}\biggr) + r_\mathrm{min}\,.
    \label{eq:inter_point_distance}
\end{equation}
The basis shape parameter is controlled by the user-defined parameter $\varepsilon_R$, which sets the value of the basis at half the distance from its nearest neighbour collocation point. In this work, $\varepsilon_R$ is consistently set to 0.9. The diameter of each basis is customarily defined as the distance within which the basis equals 0.5, hence $r_m= \sqrt{\textrm{ln}(2)}/c_m$ for the case of Gaussian bases.

\vspace{1mm}

\textbf{Step 4: Anisotropic adaptation (lines 28 to 35).} We finally reuse the smoothed gradient function $\mathbf{\tilde{G}}_{g}^\mathrm{interp}(\bm{x})$ for the anisotropic adaptation of the basis functions. The stretching is performed in the direction normal to the gradient and by an amount that increases linearly with the norm of the smoothed gradient. Similarly to works on anisotropic particle image velocimetry \citep{Novara2013} and anisotropic binning \citep{Raiola2020}, for each of the $m$ bases the aspect ratio is computed with a linear mapping $\|\mathbf{\tilde{G}}_{g}^\mathrm{interp}(\bm{x})\|_2 \in [0, 1]$ to $\text{AR} \in [1, \text{AR}_\mathrm{max}]$ defined as:
\begin{equation}
    \text{AR}_m(\mathbf{x}_{c,m}) = \|\mathbf{\tilde{G}}_{g}^\mathrm{interp}(\mathbf{x}_{c,m})\|_2  \big(\text{AR}_{\max}-1 \big) + 1\,,
\end{equation}
with $\text{AR}_\mathrm{max}$ the maximum allowed aspect ratio for the bases. The main difference with the formulation in \citet{Raiola2020} is the absence of control parameters in the stretching equation. The main control of the anisotropy is given by $\sigma_S$, responsible for the smoothing of $\mathbf{G}$ to obtain $\tilde{\mathbf{G}}$: large values of $\sigma_S$ result in small gradients, which promotes a smooth transition between isotropic and anisotropic bases, while the opposite occurs for small values of $\sigma_S$.

Finally, the stretching is carried out in a way that preserves the effective support area, thus preserving the determinant of the metric $\mathbf{M}_m$. This is achieved in two steps (lines 32 and 33): first, for every basis $m$, given the normal and tangential vectors $\mathbf{t}_m = {\mathbf{\tilde{g}}_m}/{\| \mathbf{\tilde{g}}_m \|_2}$ and $\mathbf{n}_m = [-t_{m,y}, t_{m,x}]^\T$, the stretching is computed as
\begin{equation}
\mathbf{M}_m^\text{raw} = \text{AR}_m \cdot \mathbf{t}_m \mathbf{t}_m^\T + \frac{1}{\text{AR}_m} \cdot \mathbf{n}_m \mathbf{n}_m^\T\,.
\end{equation} 

Second, introducing the scaling $s_m^2 = \det(\mathbf{M}_m^{\text{iso}})/{\det(\mathbf{M}_m^\text{raw})}$, the final metric is computed as $\mathbf{M}_m = s_m \cdot \mathbf{M}_m^\text{raw}$.

\subsection{The gradient informed formalism}\label{sec2p3}

The steps in Sec.~\ref{sec2p2} produce a basis matrix consisting of anisotropic bases tailored to the data anisotropy, but we found this to be not enough to eliminate spurious oscillations in regions of high gradients. To further mitigate oscillations, the proposed approach extends the least-squares system to fit both the function values {and their gradients} simultaneously. For a 2D problem, this extended least square problem reads

\begin{align}
A(\mathbf{w}) =\ 
&\underbrace{\frac{1}{n_P} \left\|\tilde{\mathbf{f}}(\mathbf{X}_\ast;\mathbf{w})  - \mathbf{f}_\ast \right\|_2^2}_{\text{Data fitting}} \nonumber\\
&+ \underbrace{\frac{\alpha_{g,x}}{n_{g,x}} 
  \left\| \frac{\partial \tilde{\mathbf{f}}}{\partial x} (\mathbf{X}_{g,x}) - \mathbf{g}_{g,x} \right\|_2^2}_{\text{Data gradient penalty in $x$}} \nonumber\\
&+ \underbrace{\frac{\alpha_{g,y}}{n_{g,y}} 
  \left\| \frac{\partial \tilde{\mathbf{f}}}{\partial y} (\mathbf{X}_{g,y}) - \mathbf{g}_{g,y} \right\|_2^2}_{\text{Data gradient penalty in $y$}} \nonumber\\
&+ \underbrace{\sum_{k=1}^{n_C} \frac{ \alpha_{C, k}}{n_{C,k}} 
  \left\| \left( \mathbf{\mathcal{L}}_k \tilde{\mathbf{f}}(\mathbf{X}_{C, k};\mathbf{w}) \right) \right\|_2^2}_{\text{Linear soft constraints}} \nonumber\\
&+ \underbrace{\frac{\alpha_r}{n_B} \left\| \mathbf{w} \right\|_2^2}_{\text{Tikhonov regularization}} .
\label{J_w}
\end{align} where $\tilde{\mathbf{f}}(\mathbf{X_\ast};\mathbf{w})=\mathbf{\Gamma}(\mathbf{X_\ast})\mathbf{w}$ is the RBF approximation \eqref{eq:rbf_model} of the underlying function $f$ in the set of data points $\mathbf{X_\ast}$, parametrized by the weights $\mathbf{w}$. 

The first term enforces fidelity to the $n_P$ known data values, sampled in $\mathbf{X}_\ast\in\mathbb{R}^{n_P\times 2}$ and stored in the vector \( \mathbf{f}_\ast\in\mathbb{R}^{n_P} \). The second and third terms penalize discrepancy between the predicted partial derivatives and a set of data target gradients, taken from the local polynomial regression $\mathbf{g}_{g,x}\in\mathbb{R}^{n_{g,x}}$, $\mathbf{g}_{g,y}\in\mathbb{R}^{n_{g,y}}$ described in the previous section. We highlight that these are imposed over potentially differing sets of points 
$\mathbf{X}_{g,x} \in \mathbb{R}^{n_{g,x}\times 2}$ and $\mathbf{X}_{g,y} \in \mathbb{R}^{n_{g,y}\times 2}$ and could be weighted differently, according to the weight coefficients $\alpha_{g,x},\alpha_{g,y}\in\mathbb{R}$. Moreover, the two sets of points $\mathbf{X}_{g,x}$ and $\mathbf{X}_{g,y}$ can be chosen within specific regions and activated only for points where the gradient absolute values are below a certain threshold, respectively $|\mathbf{g}_{g,x}| < \varepsilon_{g, x}$ and $|\mathbf{g}_{g,y}| < \varepsilon_{g, y}$. 

There are two main motivations for penalizing gradients only when they are small. First, the goal is to discourage oscillations, which typically occur in flat regions adjacent to strong gradients. Second, high gradients are often the result of failures in the local polynomial regression and therefore should not be given excessive weight.

\textcolor{black}{The fourth term in \eqref{J_w} implements linear penalties, that is, soft constraints. These are applied on the sets $(X_{C,1}, X_{C,2}, \ldots, X_{C,n_C})$, each containing $(n_{C,1}, n_{C,2}, \ldots, n_{C,n_C})$ points and associated with relative weights $(\alpha_{C,1}, \alpha_{C,2}, \ldots, \alpha_{C,n_C})$. This formulation provides a flexible mechanism to incorporate problem-specific constraints, such as boundary conditions or other physically motivated relations (e.g., divergence-free constraints), when relevant to the flow under consideration.}

The augmented cost function \( A(\mathbf{w}) \) \eqref{J_w} is quadratic in the coefficients \( \mathbf{w} \), and its minimization corresponds to solving a linear system. Significant gains in robustness can be achieved by combining the RBF basis, which is primarily local, with a global basis. The classic choice for the global basis, also pursued in this work, is a set of polynomial terms. Equation \eqref{eq:rbf_model} thus becomes

\begin{equation}
\label{f_P}
\tilde{f}(\bm{x})\approx \sum^{n_{B}}_{m=1} w_m \gamma_m(\bm{x})+\sum^{n_{O}}_{m=1} \beta_m p_m(\bm{x})\,,
\end{equation} 
where $n_B+n_O=n_{B^\ast}$ is the total number of basis elements. The associated matrix formulation reads:

\begin{equation}
\tilde{\mathbf{f}}(\mathbf{X}_\ast) = 
\underbrace{
\begin{pmatrix}
\mathbf{\Gamma}(\mathbf{X}_\ast) & \mathbf{P}(\mathbf{X}_\ast)
\end{pmatrix}
}_{\mathbf{A}_d}
\underbrace{
\begin{pmatrix}
\mathbf{w} \\
\mathbf{\bm{\beta}}
\end{pmatrix}
}_{\mathbf{z}},
\end{equation} where \( \mathbf{\Gamma} \in \mathbb{R}^{n_P \times n_B} \) is the RBF matrix with entries \( \gamma_m(\mathbf{x}_i) \), \( \mathbf{P}\in \mathbb{R}^{n_P \times n_O} \) is the polynomial matrix with entries \( \mathbf{P}(\mathbf{X}_\ast)[i,j]=p_j(\mathbf{x}_i) \), \( \mathbf{w} \in \mathbb{R}^{n_B} \) are the RBF weights, $\mathbf{\bm{\beta}} \in \mathbb{R}^{n_O}$ the polynomial weights, and \( \mathbf{z} \in \mathbb{R}^{n_{B^\ast}} \) the combined weights.

The linear system obtained by minimizing \eqref{J_w} with the approximation \eqref{f_P} reads:
\begin{equation}
\label{aug}
\mathbf{A}_{\text{aug}}\, \mathbf{z} = \mathbf{b}_{\text{aug}}\,,
\end{equation}
where \( \mathbf{A}_{\text{aug}} \) and \( \mathbf{b}_{\text{aug}} \) gather all the terms in the cost function and for both basis types. As an illustrative example, consider the case where gradient penalisation is applied to both the RBF and polynomial bases, a single linear soft constraint is imposed only on both bases, and Tikhonov regularisation is applied exclusively to the RBF weights. The matrix \( \mathbf{A}_{\text{aug}} \) takes the form

\begin{equation}
\mathbf{A}_{\text{aug}} =
\begin{pmatrix}
\mathbf{\Gamma}(\mathbf{X}_\ast) / \sqrt{n_P} & \mathbf{P}(\mathbf{X}_\ast)/\sqrt{n_P} \\
\sqrt{\alpha_{g,x} / n_{g,x}}\, \mathbf{\Gamma}_x(\mathbf{X}_{g,x}) & \sqrt{\alpha_{g,x} / n_{g,x}}\, \mathbf{P}_x(\mathbf{X}_{g,x}) \\
\sqrt{\alpha_{g,y} / n_{g,y}}\, \mathbf{\Gamma}_y(\mathbf{X}_{g,y}) & \sqrt{\alpha_{g,y} / n_{g,y}}\, \mathbf{P}_y(\mathbf{X}_{g,y}) \\
\sqrt{\alpha_{C,1} / n_{C,1}}\, \mathbf{\Gamma}(\mathbf{X}_{C,1}) & \sqrt{\alpha_{C,1} / n_{C,1}}\, \mathbf{P}(\mathbf{X}_{C,1}) \\
\sqrt{\alpha_r / n_{B}}\, \mathbf{I}_{n_B} & \mathbf{0}
\end{pmatrix},
\end{equation} while the associated right hand side \( \mathbf{b}_{\text{aug}} \) is
\begin{equation}
\mathbf{b}_{\text{aug}} =
\begin{pmatrix}
\mathbf{f}_\ast / \sqrt{n_P} \\
\sqrt{\alpha_{g,x} / n_{g,x}}\, \mathbf{g}_x \\
\sqrt{\alpha_{g,y} / n_{g,y}}\, \mathbf{g}_y \\
\mathbf{0}\in\mathbb{R}^{n_{C,1}} \\
\mathbf{0}\in\mathbb{R}^{n_B}
\end{pmatrix}.
\end{equation} 


The system \eqref{aug} is generally overdetermined and solved in the least-squares sense, yielding the optimal coefficients for the expansion in \eqref{f_P}. In this work, this system was solved using the standard SVD-based solver via Numpy's \texttt{lstq} routine, with a tunable tolerance parameter that discards smaller singular values and thus mitigates numerical instabilities.

\textcolor{black}{
In the present formulation, the matrices arising from the RBF regression (e.g., $\Gamma$) are dense, as typical of global RBF approaches. The memory requirement therefore scales as $\mathcal{O}(N\,n_b)$, where $n_b$ is the number of basis functions and $N$ is the total number of collocation points, including data, gradient, and constraint points. For the problem sizes considered here, $N$ ranges from approximately $5\times10^4$ to $5\times10^5$, while $n_b$ ranges from $10^3$ to $10^4$. Although this scaling remains characteristic of global formulations in the absence of localization (e.g., PUM), the proposed adaptive and anisotropic strategies significantly reduce $n_b$, thereby lowering both memory footprint and computational cost.
}

\subsection{A note on hyper-parameter tuning}

Both the algorithm for constructing the basis matrix \(\mathbf{\Gamma}(\mathbf{X}_\ast) \in \mathbb{R}^{n_P \times n_B}\) (see Sec.~\ref{sec2p2}) and the gradient-informed formalism used to solve the RBF regression problem in \eqref{J_w} (see Sec.~\ref{sec2p3}) involve a substantial number of hyperparameters. Specifically, the resampling and basis-preparation stages require parameters such as \((n_F,\, \sigma_S,\, r_{\min},\, r_{\max},\, \varepsilon_R,\, \text{AR}_{\max})\), while the gradient-informed regression introduces additional parameters \((\varepsilon_{g,x},\, \varepsilon_{g,y},\, \alpha_{g,x},\, \alpha_{g,y})\). Although this may appear daunting, reasonable values for most parameters can be readily inferred from physical intuition or basic empirical considerations. Moreover, straightforward data-driven optimization techniques for hyperparameter selection are available and have also been tested in this work.

The number of points used in the local regression, \(n_F\) in \eqref{eq:least_squares_polyfit}, is selected as a compromise between noise reduction and the ability to capture local gradient variations accurately. Its value can be estimated from the available sample density and the characteristic spatial scale over which the regression is intended to remain ``local.'' The smoothing parameter \(\sigma_S\) controls the level of anisotropic sampling and collocation, and therefore has the largest impact on the overall methodology. The parameters \(r_{\min}\) and \(r_{\max}\) define the minimum and maximum collocation densities, and suitable ranges for their values can be inferred from the expected gradient magnitudes. The parameter \(\varepsilon_R\), first introduced in \citet{Sperotto2022a}, defines the RBF size before stretching; the default value \(\varepsilon_R=0.9\) is used throughout this work. Finally, \(\text{AR}_{\max}\) sets the maximum stretching of the basis. A value in the range 4–-6 provides a good compromise between anisotropy and numerical stability, consistent with previous findings on anisotropic binning and windowing \citep{Novara2013, Raiola2020}.

Less intuitive is the selection of the hyper-parameters associated with the gradient-informed regression, since their optimal values strongly depend on the quality of the gradient estimates. When this is affected by unresolved regions or local inaccuracies, enforcing excessive compliance with poorly resolved gradients may degrade the overall reconstruction quality. This is particularly challenging because the reliability of the estimated gradients is generally not known a priori. Nevertheless, this type of dilemma is well known in physics-informed regression methods ---including the class of PINNs---which must balance empirical data fidelity against physically motivated loss or constraint terms \citep{Rohrhofer2023}. 

To address this optimal-weighting challenge in a principled manner, we employ a structured hyperparameter optimization framework. \textcolor{black}{For each test case, the available dataset is partitioned into two disjoint subsets: a training set, used to compute the RBF weights, and a testing set, used for hyperparameter selection. Both subsets are extracted from the same dataset using identical criteria, ensuring statistical consistency.} The weights are obtained through regression on the training data, while the hyperparameter optimization is carried out by minimizing the reconstruction error evaluated on the testing data using the Optuna framework \citep{optuna_2019}. This procedure promotes the selection of hyperparameters that generalize beyond the data used for fitting and mitigates overfitting.

\section{Selected Test Cases}\label{sec.test}

Two test cases are considered to evaluate the performance of the proposed approach against a traditional isotropic formulation. The first test case, described in Sec.~\ref{sec3p2}, is built from the well-known turbulent channel flow based on direct numerical simulations (DNS). This provides the ground truth to test the method performance in terms of gradient resolution and statistical convergence. The second, described in Sec.~\ref{sec3p3}, is the 3D tracking velocimetry of a turbulent jet flow, from which only a qualitative assessment is possible. In both cases, isotropic and anisotropic regressions are carried out using identical solver parameters to ensure a fair comparison. Both datasets are made available at \url{https://github.com/mendezVKI/SPICY_VKI}.

\subsection{DNS of a channel flow}\label{sec3p2}

We consider the turbulent channel flow at \(\text{Re}_\tau \approx 1000\) from the Johns Hopkins Turbulence Database \citep{Li2008, Graham2016, Graham2017}. The simulation domain spans \([8\pi h \times 2h \times 3\pi h]\), where \(h\) denotes the channel half-height, and is discretized over \(2048 \times 512 \times 1536\) grid points in the streamwise, wall-normal, and spanwise directions, respectively. The dataset covers a dimensionless time interval \(t U_b / h \in [0,\, 25.9935]\), where \(U_b\) is the bulk velocity. The DNS time step is \(\delta t U_b/h = 0.0013\), and data are available every \(\Delta t = 5\delta t\), yielding \num{4000} recorded snapshots. In this study, we focus on the streamwise and wall-normal velocity components, \(u\) and \(v\).

\begin{figure}
    \centering
    \includegraphics[width=0.49\textwidth]{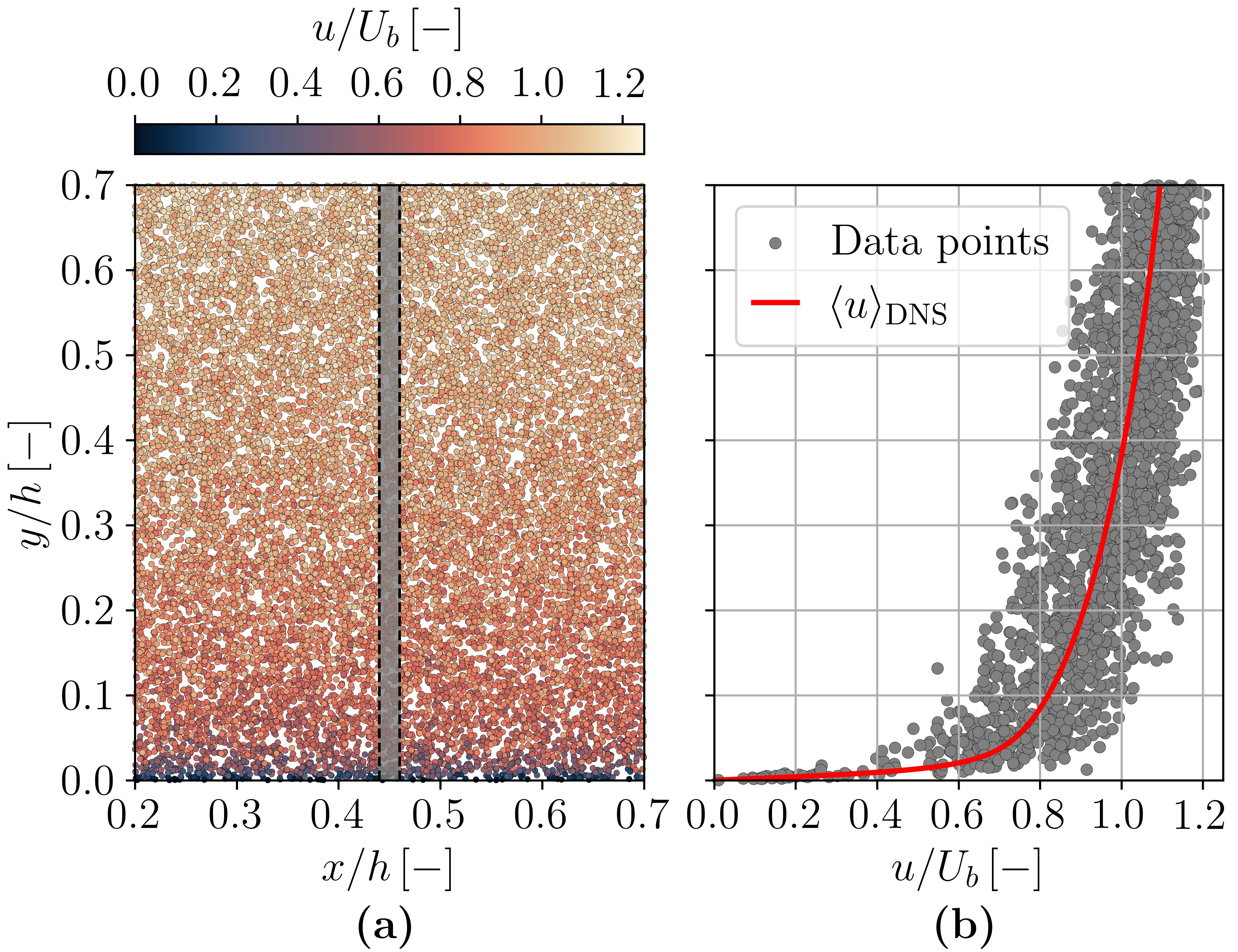}
    \caption{Test case 1: (a) Scatter plot of the velocity ensemble for the regression of average velocity, coloured by streamwise velocity component, with the vertical slice region at $x/h=0.45$ highlighted in gray. (b) Data points extracted from the highlighted slice with the time-average DNS velocity profile (solid, red line)}
    \label{fig:case2_data_points}
\end{figure}

The data are sampled in two stages. The first stage follows the approach proposed in \citet{Raiola2020}, consisting of random sampling in the \(x\)–\(y\) plane, while the \(z\) and \(t\) axes are discretized over \num{6000} frames. These frames are obtained by combining 30 equally spaced positions in \(z/h \in [0,\, 3\pi]\) with 200 equally spaced time instants in \(t U_b/h \in [0,\, 25.9935]\). For each frame, 1000 spatial points are randomly selected within the domain \((x/h, y/h) \in [0,\, 2] \times [0,\, 1]\), resulting in a total of \num{6000000} sampled points. Spatial interpolation from the DNS grid to the randomly sampled locations is performed using 5th order splines computed over 8 data points \citep{Graham2016}, while no temporal interpolation is used since the selected time steps coincide with those of the original simulation. 

While the total number of $n_P^{\text{full}} = \num{6000000}$ can be seen as the set of data collected in an experiment, the regression was carried out in a subsampled set of $n_P=\num{50000}$ for the mean flow and $n_P=\num{500000}$ for the fluctuating component. For this test case, both the case of uniform sampling and adaptive sampling were considered.

Figure \ref{fig:case2_data_points}(a) shows a scatter plot using uniform sub-sampling with $n_P = \num{50000}$, with markers coloured by the value of the streamwise velocity component. A slice in $x/h\in[0.44,0.46]$ is highlighted and the samples in this slice are shown in Fig.~\ref{fig:case2_data_points}(b), to highlight the main challenge of this test case: in addition to the sharp gradient near the wall, the ensemble features a significant variance due to turbulence. The DNS profile of the average streamwise velocity component is drawn with a solid, red line.

\subsection{3D velocimetry of a turbulent round jet}\label{sec3p3}

\begin{figure}
    \centering
    \includegraphics[width=0.49\textwidth]{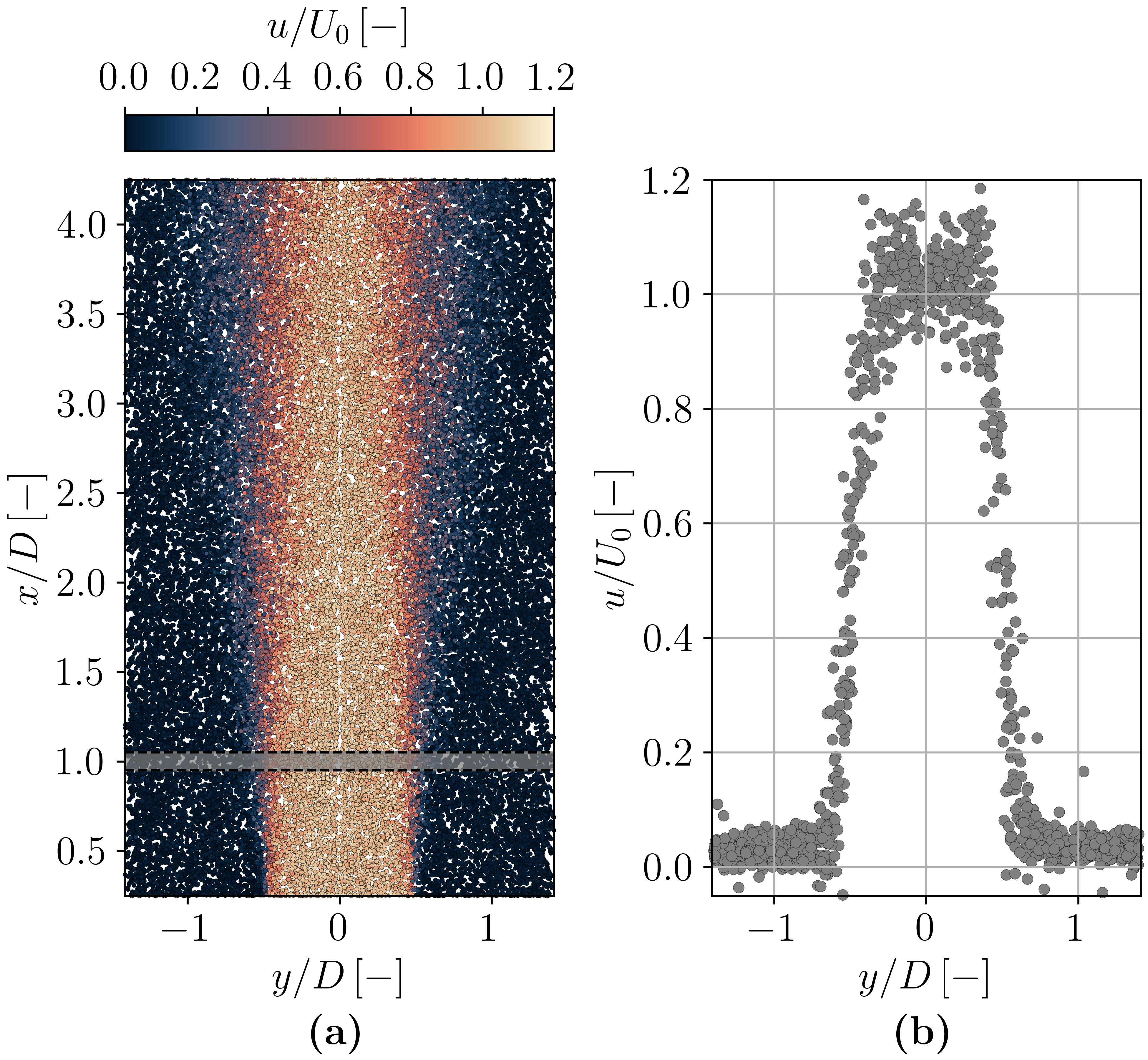}
    \caption{Test case 2: (a) Scatter plot of the velocity ensemble for the regression of average velocity, coloured by streamwise velocity component, with the horizontal slice region at $x/h=1$ highlighted in gray. (b) Data points extracted from the highlighted slice}
    \label{fig:case3_data_points}
\end{figure}
The second test case considers a turbulent water jet measured using 3D PTV. 
The coordinate system is defined such that $y$ denotes the radial direction, 
$x$ the axial direction along the jet, and $z$ the direction perpendicular to both. 
The jet issues upward in a hexagonal acrylic tank, yielding a 
diameter-based Reynolds number of $\mathrm{Re}=6750$, with nozzle diameter 
$D=50\ \mathrm{mm}$. The measurement volume spans 
$x/D \times y/D \times z/D \in [0.25, 4.25] \times [-1.3, 1.3] \times [-0.5, 0.5]$.
Four cameras acquired \num{2000} time-resolved snapshots at 
\SI{1000}{\hertz}. Further details of the experimental setup 
are reported by \citet{Ratz2024}.

To perform a two-dimensional regression, the 3D point cloud is reduced to a single $x$--$y$ plane by grouping all measurements according to their radial distance from the jet centreline and treating points with the same radius as belonging to the same plane. \textcolor{black}{This operation exploits the well-established statistical axisymmetry of turbulent round jets, for which the mean flow and second-order statistics primarily depend on the axial and radial coordinates. Under this assumption, points at the same radial distance can be treated as statistically equivalent, enabling a consistent and more robust regression.}
\textcolor{black}{
On the other hand, this projection significantly alters the uniformity of the sampling: under} the assumption of approximately uniform sampling in 3D, this projection produces a two-dimensional ensemble whose point density increases proportionally to $y^2$. The resulting density imbalance is compensated for using the adaptive downsampling strategy introduced in Sec.~\ref{sec.methodology}. After resampling, $n_P = \num{50000}$ points are retained for the regression of the mean velocity component ($u$), and $n_P = \num{500000}$ points for the fluctuating component ($u'$). The spatial distribution of the resampled points for the streamwise component is displayed in Fig.~\ref{fig:case3_data_points}(a), while Fig.~\ref{fig:case3_data_points}(b) 
shows a slice in the region $x/D \in [0.95, 1.05]$, highlighting the intensity of the gradients and the local turbulence levels.

\section{Results}\label{sec.results}

On both selected test cases, three different configurations of the RBF regression are examined, and described below. For convenience, short identifiers for each configuration are given in parentheses.

\begin{enumerate}
\item \textbf{Traditional RBFs with uniform collocation (IsoUni).} Uniformly distributed, isotropic RBF regression without gradient penalisation
(i.e.\ $\alpha_{g,x}=\alpha_{g,y}=0$). \textcolor{black}{The best performance of this approach is obtained by selecting the number of bases $n_b$ that provides the lowest reconstruction error.}

\item \textbf{Isotropic RBFs with adaptive collocation (IsoAdapt).} \textcolor{black}{This configuration uses the same isotropic bases as IsoUni, but with adaptive collocation, and still without anisotropic adaptation or gradient penalisation.}

\item \textbf{Proposed approach (AnisoAdapt).} This is the full methodology described in
section~\ref{sec.methodology}: Adaptive collocation combined with anisotropic adaptation and gradient penalisation. \textcolor{black}{In this case, the number of bases is reduced by a factor of five compared to the previous methods, as anisotropic bases can represent flow structures more efficiently, which significantly reduces the computational cost of the regression.}
\end{enumerate}

\begin{table*}[t]
    \centering
    \begin{tabularx}{\textwidth}{
        l
        *{4}{>{\centering\arraybackslash}X}
        *{3}{>{\centering\arraybackslash}X}
    }
        \toprule
        & \multicolumn{4}{c}{Test case 1} 
        & \multicolumn{2}{c}{Test case 2} \\
        \cmidrule(lr){2-5}
        \cmidrule(lr){6-7}
        Collocation method 
            & $\langle u \rangle$ 
            & $\langle v \rangle$ 
            & $\langle u'u' \rangle$ 
            & $\langle v'v' \rangle$
            & $\langle u \rangle$ 
            & $\langle u'u' \rangle$
            \\
        \midrule
        Isotropic, uniform 
            & $\bm{186}$ & $\bm{3}$ & $\bm{1693}$ & $\bm{1756}$
            & $\bm{214}$ & $\bm{3026}$ \\
         - gradients est. and adapt. samp. 
            & 15 & - & 1064 & 896
            & 152 & 613 \\
         - uniform collocation 
            & 1 & 1 & 1 & 1
            & 1 & 1 \\
         - regression 
            & 170 & 2 & 628 & 859
            & 61 & 2412 \\
         MSE
            & $1.3e-4$ &  & $9.8e-7$ & $1.3e-8$
            & - & - \\[3pt]

        Isotropic, adaptive 
            & $\bm{32}$ & \textbf{-} & $\bm{1108}$ & $\bm{957}$
            & $\bm{157}$  & $\bm{711}$ \\
         - gradients est. and adapt. samp. 
            & 15 & & 1064 & 896
            & 152 & 613 \\
         - adaptive collocation 
            & 11 & & 4 & 2
            & 1 & 2 \\
         - regression 
            & 6 & & 40 & 59
            & 4 & 96 \\
         MSE
            & $3.3e-5$ &  & $4.6e-7$ & $5.9e-9$
            & - & - \\[3pt]

        Anisotropic, adaptive 
            & $\bm{36}$ & \textbf{-} & $\bm{1120}$ & $\bm{968}$
            & $\bm{166}$ & $\bm{738}$ \\
         - gradients est. and adapt. samp. 
            & 15 & & 1064 & 896
            & 152 & 613 \\
         - adaptive collocation 
            & 11 & & 2 & 2
            & 1 & 2 \\
         - stretching 
            & 1 & & 1 & 1
            & 1 & 1 \\
         - regression 
            & 9 & & 53 & 69
            & 12 & 122 \\
         - MSE
            & $1.2e-5$ &  & $2.9e-8$ & $9.3e-10$
            & - & - \\
        \bottomrule
    \end{tabularx}

    \caption{CPU times (in seconds) for all regression approaches for test case 1 and test case 2. \textcolor{black}{For each approach, the Mean Square Error (MSE) with respect to the ground truth is also reported.}}
    \label{tab:computation_times}
\end{table*}

This comparative analysis makes it possible to isolate the individual effects of adaptive collocation, anisotropic adaptation, and gradient-informed regularisation, thereby clarifying the contribution of each component to the overall reconstruction quality.

Since the IsoUni approach typically requires a substantially larger number of basis functions to adequately cover the domain, its computational cost is significantly higher. In contrast, adaptive collocation concentrates basis centres only where they are needed, resulting in a markedly more efficient regression. The computation times for both test cases are
reported in Table~\ref{tab:computation_times}. \textcolor{black}{A comparison of the mean squared error (MSE) between the regression and the DNS profile is provided as a more quantitative comparison.} All computations were performed on a standard laptop equipped with an AMD Ryzen~7~PRO~6850U processor (2.70\,GHz) and 32\,GB of RAM.

\subsection{DNS of a channel flow}
This section details the regression results for test case 1, beginning with the mean velocity components $\langle u \rangle$ and $\langle v \rangle$, followed by the analysis of the velocity fluctuations $\langle u'u' \rangle$ and  $\langle v'v' \rangle$.

For the average streamwise velocity $\langle u \rangle$, both the uniform and adaptive downsampling strategy are investigated, resulting in a downsampled dataset $\mathbf{X}_*$ of $n_P = \num{50000}$ points, selected within the subdomain $(x/h, y/h) \in [0.2, 0.7] \times [0, 0.7]$. All data are aggregated along the $z$ and $t$ dimensions into a single $x$–$y$ frame as presented in the previous section. This downsampling step significantly reduces computational cost with no appreciable loss of accuracy.

Starting with the uniform sampling, the collocation of IsoUni is performed using a constant radius of $r_c \approx 0.005$, such to obtain  $n_B = \num{7975}$ from the Poisson disk algorithm. The bases are shaped using $\varepsilon_R=0.9$. The same parameters are kept for the following approaches. For the IsoAdapt approach, the gradient estimates used to adapt the collocation points are computed on $\mathbf{X}_g$ with $n_G=n_P/10=\num{5000}$, randomly taken from $\mathbf{X}_*$, and the local fitting is achieved with $n_F=200$ data points. 

\begin{figure}
    \centering
    \includegraphics[width=1\linewidth]{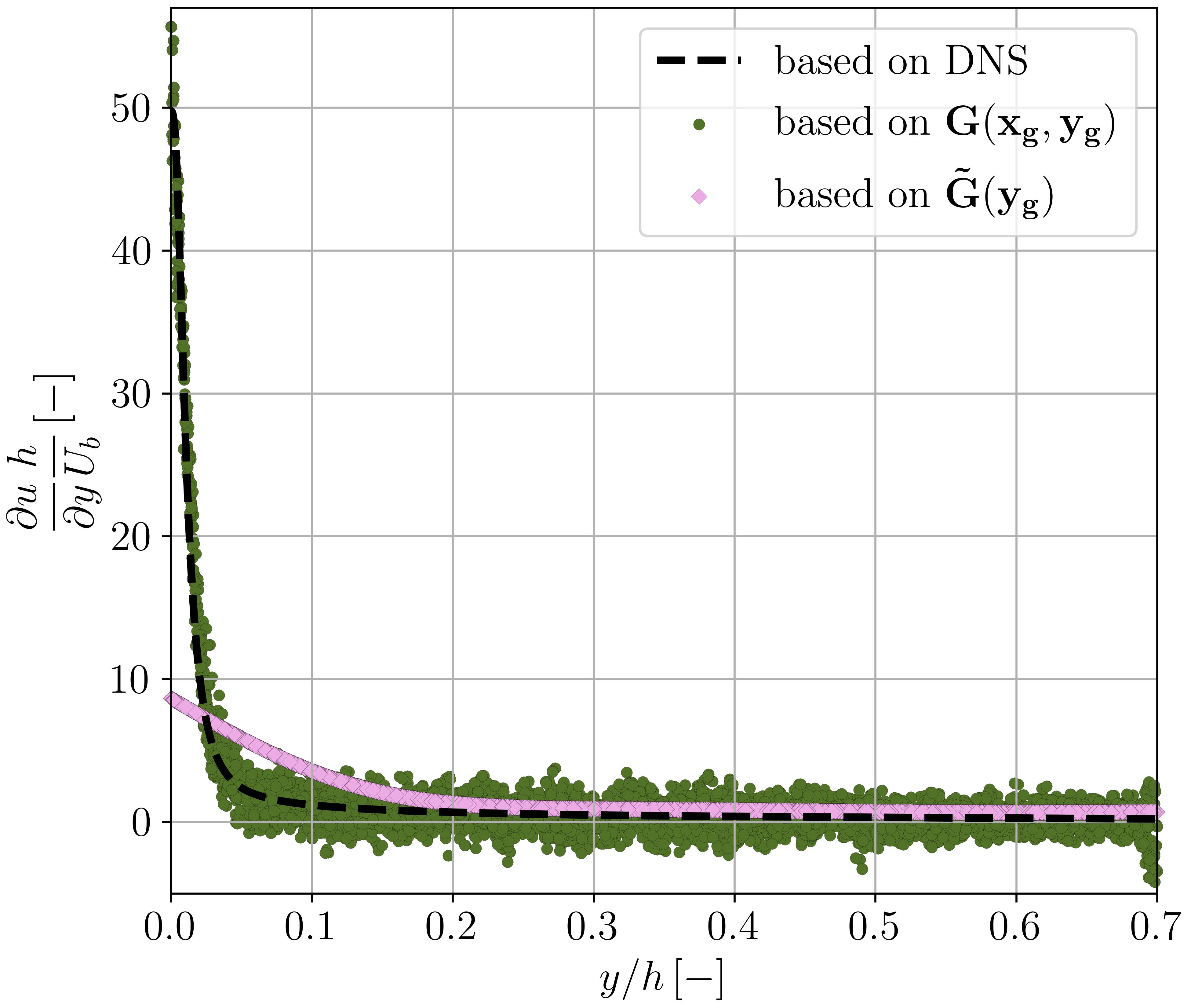}
    \caption{Comparison between the ground truth gradient from the DNS, the estimated gradient and the gradient smoothed with $\sigma_S=0.05$}
    \label{fig:case2_grad_smoothing}
\end{figure}

Leveraging the full development of the flow, the gradient along the streamwise direction is set to zero, i.e. $\mathbf{g}_{g,x}=\tilde{\mathbf{g}}_{g,x}=\bm{0}$. Moreover, the cross streamwise gradient field is estimated from the ensemble-averaged dataset obtained by collapsing all samples into a single profile, which is subsequently smoothed with a Gaussian kernel of width $\sigma_s \approx 0.075$. This smoothing enforces a fully developed gradient field and promotes a uniform distribution of basis functions along $x$. A comparison between the raw and smoothed gradients is presented in Fig.~\ref{fig:case2_grad_smoothing}. The smoothing is deliberately aggressive: if the collocation was based on more accurate gradients, the resulting bases would exhibit excessively sharp variations in their shape factors, which could compromise the conditioning of the regression system.

\begin{figure}
    \centering
    \includegraphics[width=1\linewidth]{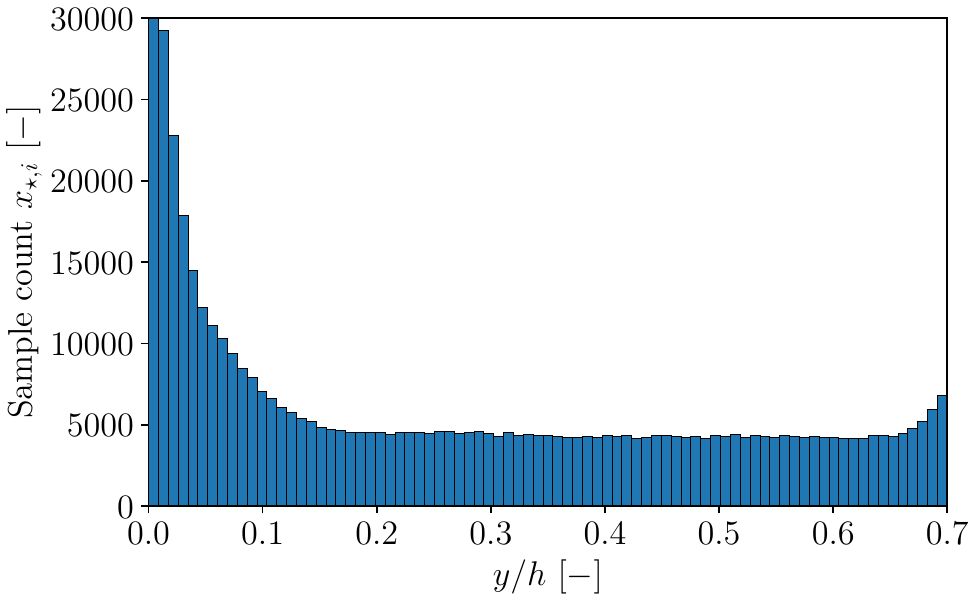}
    \caption{Test case 1: Number of data samples along each position $y/h$ from the adaptive downsampling approach}
    \label{fig:adapt_sampling}
\end{figure}

\begin{figure*}
    \centering
    \includegraphics[width=0.98\textwidth]{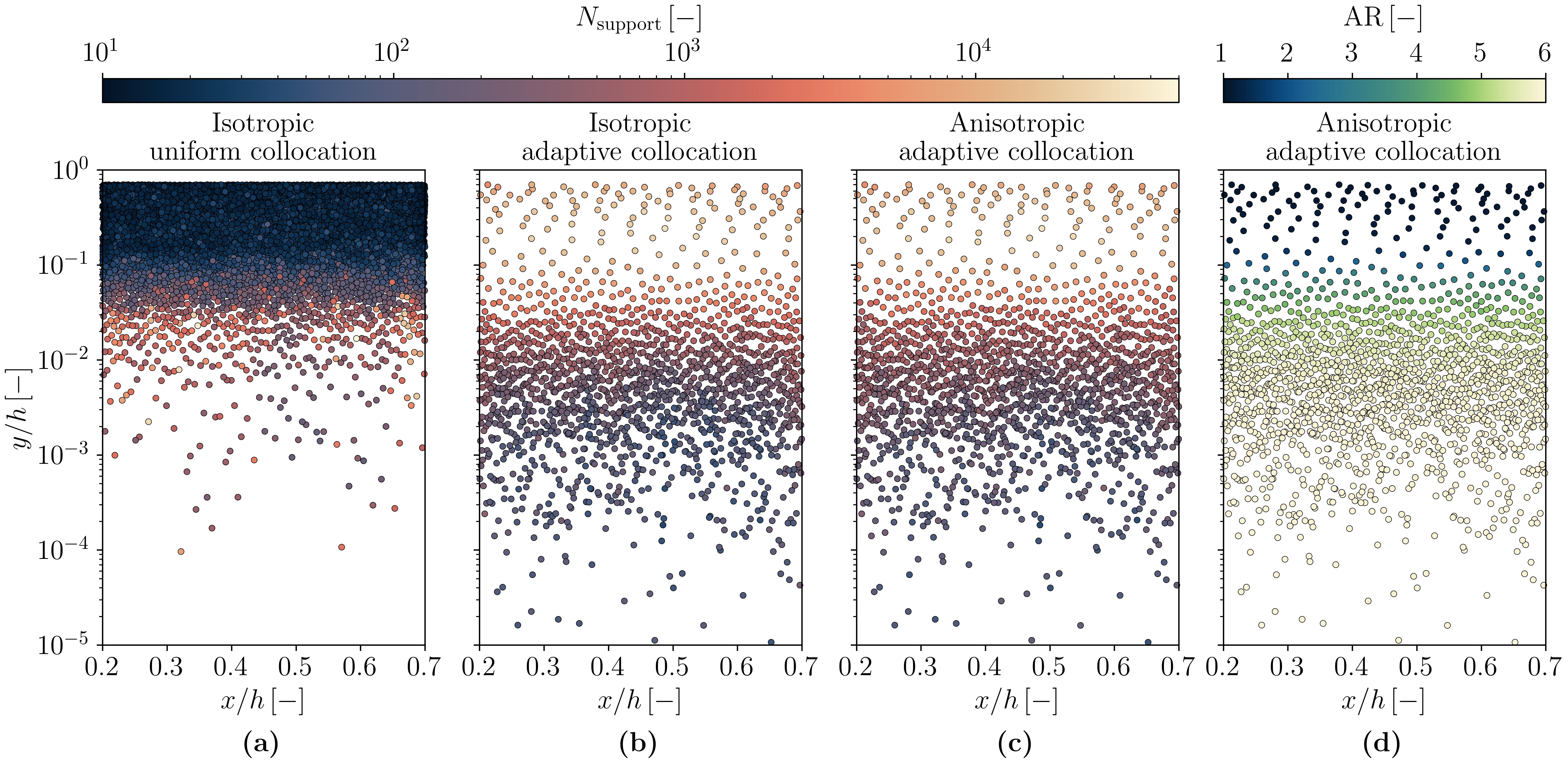}
    \caption{Test case 1: Spatial locations of the collocation points, coloured by the number of supporting points using the adaptive data sampling for the isotropic uniform (a), isotropic adaptive (b) and anisotropic adaptive basis (c). Aspect ratio (AR) of the anisotropic basis elements (d)}
    \label{fig:case2_point_support}
\end{figure*}

Using $\tilde{\mathbf{G}}_g = [\,\bm{0} \;\; \tilde{\mathbf{g}}_{g,y}(y) \, ]$, the adaptive collocation is performed with a variable radius $r_c \in [0.001,\, 0.05]$, resulting in $n_B = \num{1595}$ basis centers. We emphasize that, compared to the IsoUni approach, the number of bases is reduced by a factor of 5.

The AnisoAdapt approach uses the same collocation but is complemented by the stretching of the bases, which is performed according to $\tilde{\mathbf{G}}_g^{\text{interp}}$, and up to $\text{AR}_\mathrm{max}=6$.

For all three approaches, a wall no-slip condition is imposed on a set of points $\mathbf{X}_{C,1}$ of size $n_{C,1} = 1000$, located at $y = 0$ and uniformly distributed over the range $x \in [0.2, 0.7]$. The systems are then solved using $\alpha_r=0.02$, $\alpha_{c,1}=100$, for all approaches. The anisotropic method involves gradient penalisation parameters to mitigate the oscillations. A penalisation of the x gradients is performed on a set of points $\mathbf{X}_{g,x}$. No penalty is applied in the y-direction, as the $\partial u / \partial x = 0$ condition proves significantly more effective in suppressing oscillations. The set $\mathbf{X}_{g,x} \in \mathbb{R}^{2 \times n_{g,x}}$, with $n_{g,x}=5000$ covers the entire domain ($\varepsilon_{g,x} = 10$) while $\mathbf{X}_{g,y}$ is empty ($\varepsilon_{g,y} = 0$). The optimal value of $\alpha_{g,x} = $ 4E$-$3 is found through optimisation, while $\alpha_{g,y}$ is null in this case.

The same procedure is then repeated with the adaptive downsampling, as described in Step 2 of section ~\ref{sec.methodology}. The first estimation of the gradient is performed, but this time data are re-sampled from the complete set according to a probability density function obtained from \eqref{eq:final_pdf}. This results in a set of points $\mathbf{X_*}$ with a non-uniform density distribution, as illustrated in Fig.~\ref{fig:adapt_sampling}, presenting the samples counts along the y direction. Based on this, $\mathbf{X}_{g}$ is resampled, $\mathbf{G}_{g}$ is evaluated using $n_{F}=200$, and the resulting field is smoothed with $\sigma_{S}=0.05$. The uniform collocation for the IsoUni approach is obtained with $r_c\approx0.005$, resulting in $n_B=\num{7985}$ bases. The adaptive collocation for the IsoAdapt approach uses $r_c \in [0.001, 0.05]$, which results in only $\num{1597}$ bases, while improving the effective support of each basis. This is illustrated in Figs.~\ref{fig:case2_point_support}(a)–(c), which show the spatial distribution of collocation points, coloured by the number of data samples supporting each basis element for the adaptive data sampling approach. In the traditional case (Fig.~\ref{fig:case2_point_support}(a)), the large bases density leads to a very small support for $y/h > 10^{-1}$. Reducing the global number of bases could alleviate this, but would also degrade the representation in the near-wall region ($y/h < 10^{-1}$). In contrast, the adaptive collocation employed in the IsoAdapt approach (Fig.~\ref{fig:case2_point_support}(b)) produces a more balanced basis distribution across the domain. Near the wall, the least populated bases still retain on the order of $100$ supporting points, while in the free-stream region the support increases up to $N_{\mathrm{support}} \sim \num{10000}$, ensuring both numerical stability and smoothness of the reconstruction. Finally, the AnisoAdapt approach stretches the bases up to $\text{AR}_{\text{max}}=6$. The advantage here is that the shape of the bases in the gradient direction is reduced while the number of supporting points is not changed, as shown in Fig.~\ref{fig:case2_point_support}(c). The distribution of the aspect ratio is presented in Fig.~\ref{fig:case2_point_support}(d), with a maximum stretching at the wall where the gradient is the largest, smoothly transitioning to $\text{AR}=1$, that is, uniform bases as we move away from the wall and the flow gradients become similar. For this approach, optimisation yields $\alpha_{g,x}=6$E$-3$ while all other parameters remain identical and are summarised in Table~\ref{tab:reg_param_case1}.

\begin{figure*}
    \centering
    \includegraphics[width=0.98\textwidth]{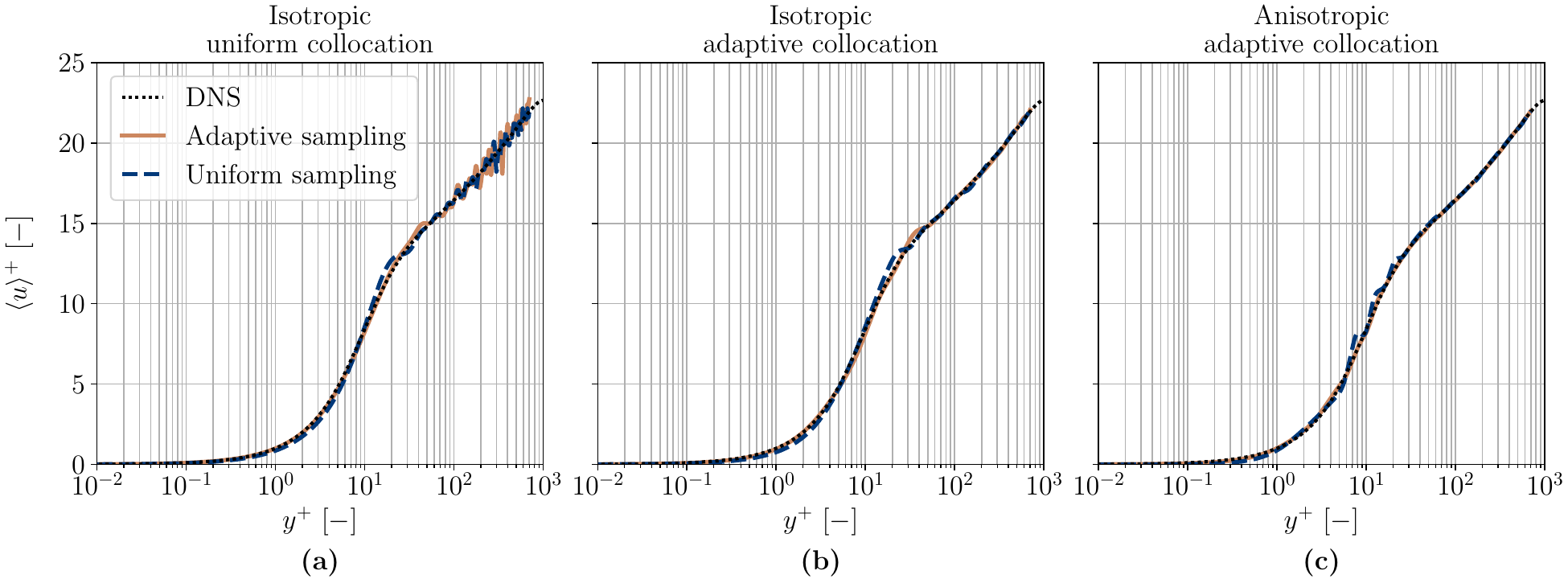}
    \caption{Test case 1: Mean velocity profiles $\langle u \rangle$ at $x/h = 0.45$ for uniform collocation (a), adaptive isotropic collocation (b) and adaptive anisotropic collocation (c). The legend in subfigure (a) applies to all subfigures}
    \label{fig:case2_u_results}
\end{figure*}

\begin{figure*}
    \centering
    \includegraphics[width=0.98\textwidth]{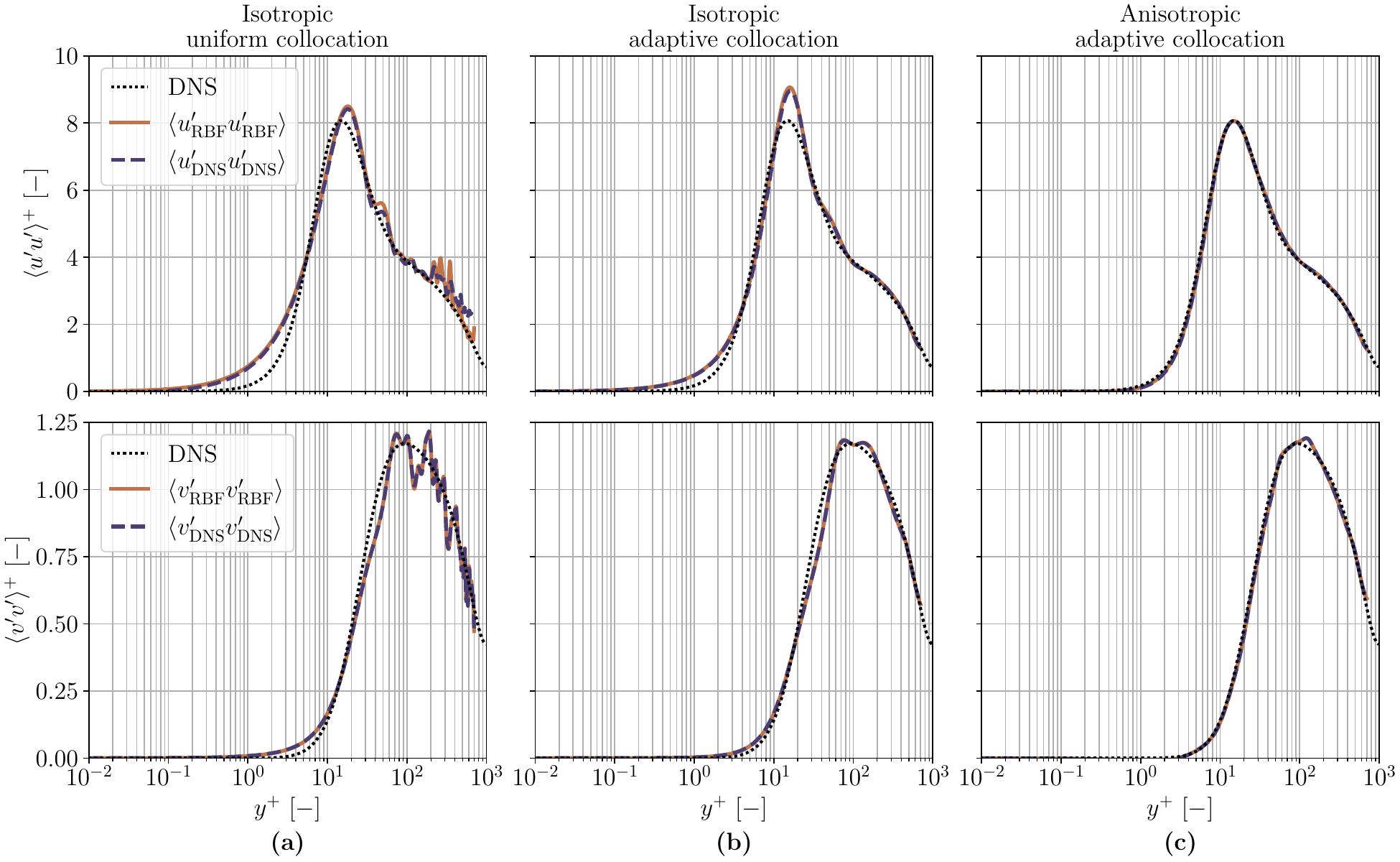}
    \caption{Test case 1: Reynolds stress profiles $\langle u^\prime u^\prime \rangle^+$ (top) and $\langle v^\prime v^\prime \rangle^+$ (bottom) at $x/h = 0.45$ for uniform collocation (a), adaptive isotropic collocation (b) and adaptive anisotropic collocation (c). The legend in subfigure (a) applies to all subfigures}
    \label{fig:case2_uu_results}
\end{figure*}

The regression for each of the 3 approaches is presented in Fig.~\ref{fig:case2_u_results}, comparing a slice of the regression at $x=0.45$, with and without the adaptive downsampling, against the DNS ground truth. For the IsoUni approach (Fig.~\ref{fig:case2_u_results}(a)), which involves a very large number of bases, an overshoot is observed at $y^+ \approx 15$, accompanied by significant oscillations for $y^+ > 100$. Decreasing the number of bases would help reducing the oscillations in this region but would also degrade the result for $y^+ <100$. Added to this is the significant computational cost of such a regression (\SI{170}{\second} for the IsoUni approach cf. Table \ref{tab:computation_times}). The adaptive downsampling of the data helps in better retrieving the gradient by reducing the overshoot, but amplifies the oscillation for $y+ > 100$.

\begin{table}[t]
    \centering
    \begin{tabularx}{\linewidth}{l *{2}{>{\centering\arraybackslash}X} *{2}{>{\centering\arraybackslash}X}}
        \hline \\[-8pt]
        & $\langle \mathbf{u} \rangle$ & $\langle \mathbf{v} \rangle$ & $\langle \mathbf{u'u'} \rangle$ & $\langle \mathbf{v'v'} \rangle$\\[2pt]
        \hline \\[-7pt]
        \textit{Gradient estimation} & & & & \\
        \hspace{\tabind} $n_P$ & \multicolumn{2}{c}{$\num{50000}$} & \multicolumn{2}{c}{$\num{500000}$} \\
        \hspace{\tabind} $n_G$ & $\num{5000}$ & - & \multicolumn{2}{c}{$\num{5000}$} \\
        \hspace{\tabind} $n_F$ & $200$ & - & \multicolumn{2}{c}{$\num{10000}$} \\
        \hspace{\tabind} $\sigma_S$ & $0.05$ & - & $0.05$ & $0.05$ \\[4pt]
        \textit{Basis definition} & & & & \\
        \hspace{\tabind} $\varepsilon_R$ & \multicolumn{2}{c}{$0.9$} & \multicolumn{2}{c}{$0.9$} \\
        \hspace{\tabind} $r_\mathrm{min}$ & 1E$-$3 & $\approx0.02$ & 2E$-$3 & 3E$-$3 \\
        \hspace{\tabind} $r_\mathrm{max}$ & $0.05$ & $\approx0.02$ & $0.05$ & $0.05$ \\
        \hspace{\tabind} $\text{AR}_\mathrm{max}$ & $6$ & $1$ & \multicolumn{2}{c}{$6$} \\[4pt]
        \textit{Fixed penalties} & & & & \\
        \hspace{\tabind} $\alpha_r$ & \multicolumn{2}{c}{$0.01$} & \multicolumn{2}{c}{$0.01$} \\
        \hspace{\tabind} $\alpha_{C,1}$ & \multicolumn{2}{c}{$100$} & \multicolumn{2}{c}{$100$} \\[4pt]
        \textit{Optimized penalties} & & & & \\
        \hspace{\tabind} $\varepsilon_{g, x}$ & \multicolumn{2}{c}{$10$} & \multicolumn{2}{c}{$10$} \\
        \hspace{\tabind} $\varepsilon_{g, y}$ & $0$ & $10$ & \multicolumn{2}{c}{$0$} \\
        \hspace{\tabind} $\alpha_{g, x}$ & 6E$-$3 & $0.02$ & 4E$-$4 & 8E$-$4 \\
        \hspace{\tabind} $\alpha_{g, y}$ & $0$ & $0.06$ & \multicolumn{2}{c}{$0$} \\[3pt]
        \hline
    \end{tabularx}
    \caption{Test case 1: Summary of the regression parameters.}
    \label{tab:reg_param_case1}
\end{table}

By comparison, the IsoAdapt approach uses only a fraction of the number of basis functions. Its main advantage is the absence of oscillations for $y^+ > 100$, although it does not eliminate the overshoot. The adaptive downsampling further improves gradient capture, and while the overshoot is significantly reduced, it is still noticeable around $y^+ \approx 40$. Finally, the adaptive anisotropic method, due to its stretching, has improved capabilities in terms of gradient resolution, avoiding typical overshoot. This increase in resolution is stabilised by the penalty term to avoid oscillations. As visible from Fig.~\ref{fig:case2_u_results}(c), the regression with the adaptive downsampling is excellent both in the close-wall region, where the gradient is well resolved without any overshoot, and in the far-wall region, where no oscillations are observed. It is worth noting that uniform sampling fails to produce satisfactory results in this case, as the number of points supporting the bases becomes insufficient, promoting convergence issues.

\begin{table}[t]
    \centering
    \begin{tabularx}{\linewidth}{l *{2}{>{\centering\arraybackslash}X} *{2}{>{\centering\arraybackslash}X}}
        \hline \\[-8pt]
        & $\langle \mathbf{u} \rangle$ & $\langle \mathbf{u'u'} \rangle$ \\[2pt]
        \hline \\[-7pt]
        \textit{Gradient estimation} & & \\
        \hspace{\tabind} $n_P$ & $\num{50000}$ & $\num{500000}$ \\
        \hspace{\tabind} $n_G$ & $\num{25000}$ & $\num{25000}$ \\
        \hspace{\tabind} $n_F$ & $\num{1000}$ & $\num{1000}$ \\
        \hspace{\tabind} $\sigma_S$ & $0.05$ & $0.03$ \\[4pt]
        \textit{Basis definition} & & \\
        \hspace{\tabind} $\varepsilon_R$ & $0.9$ & $0.9$ \\
        \hspace{\tabind} $r_\mathrm{min}$ & 8E$-$3 & 3E$-$3 \\
        \hspace{\tabind} $r_\mathrm{max}$ & $0.05$ & $0.05$ \\
        \hspace{\tabind} $\text{AR}_\mathrm{max}$ & $4$ &$4$ \\[4pt]
        \textit{Fixed penalties} & & \\
        \hspace{\tabind} $\alpha_r$ & $0.01$ &$0.01$ \\[4pt]
        \textit{Optimized penalties} & & \\
        \hspace{\tabind} $\varepsilon_{g, x}$ & $0.1$ & 0.5 \\
        \hspace{\tabind} $\varepsilon_{g, y}$ & $18.7$ & 1.5 \\
        \hspace{\tabind} $\alpha_{g, x}$ & 1E$-$3 & 3E$-$5\\
        \hspace{\tabind} $\alpha_{g, y}$ & 5E$-$6 & 3E$-$5\\[3pt]
        \hline
    \end{tabularx}
    \caption{Test case 2: Summary of the regression parameters}
    \label{tab:reg_param_case2}
\end{table}

For the cross-stream velocity component $v$, a uniform sampling procedure is used. In this case, the collocation is more straightforward, as no significant gradients are expected in the data. Consequently, a total of $n_B = 501$ uniformly distributed isotropic bases are employed, so as to obtain a ratio point-per-basis of $\approx 100$. The bases are shaped with $\varepsilon_R = 0.9$. Although no gradient estimation is required, including penalties still enhances the regression performance. The gradient penalisation terms are set to $\mathbf{g}_{g,x} = \mathbf{0} \in \mathbb{R}^{n_{g,x}}$ and $\mathbf{g}_{g,y} = \mathbf{0} \in \mathbb{R}^{n_{g,y}}$, applied everywhere with $\varepsilon_{g,x}=10$, $\varepsilon_{g,y}=10$, while the corresponding penalisation weights are optimised, yielding $\alpha_{g,x} = 0.02$ and $\alpha_{g,y} = 0.06$. The complete set of parameters is listed in Table~\ref{tab:reg_param_case1}.

The second step focuses on the regression of the velocity fluctuations, which typically requires an order of magnitude more data points than the mean flow regression. For this purpose, the adaptive downsampling strategy is required to maintain a reasonable computation cost.

We begin by computing the instantaneous velocity components $\mathbf{u}$ and $\mathbf{v}$. Using the analytical expressions of the mean velocity fields $\langle u \rangle (\bm{x})$ and $\langle v \rangle (\bm{x})$ obtained from the previous step, the velocity fluctuations $\mathbf{u'}_*$ and $\mathbf{v'}_*$ at the points $\mathbf{X_*}$ are estimated as:
\begin{equation}
\begin{aligned}
    \mathbf{u}'_* &= \mathbf{u}_* - \langle u \rangle(\mathbf{X}_*) \\
    \mathbf{v}'_* &= \mathbf{v}_* - \langle v \rangle(\mathbf{X}_*)
\end{aligned}
\label{eq:u_prime}
\end{equation}
The regression is performed on the squared fluctuation components, $\mathbf{u'_*u'_*} $ and $\mathbf{v'_*v'_*} $.

For both $\langle u'u' \rangle$ and $\langle v'v' \rangle$, we start by uniformly sampling $n_P = \num{500000}$ data points from which an estimate of the gradient is computed on $n_G = n_P/100 = \num{5000}$ data points, using $n_F = \num{10000}$. As for the regression of the mean $\langle u \rangle$, the streamwise gradient $\mathbf{g}_{g,x}$ is set to zero while the cross-stream gradient $\mathbf{g}_{g,y}$ is grouped along the $x$-axis and smoothed using $\sigma_S=0.05$. Based on the gradient estimation $\tilde{\mathbf{G}}_g = [\,\bm{0} \;\; \tilde{\mathbf{g}}_{g,y} \, ]$, the adaptive downsampling is performed, this time using $||\tilde{\mathbf{G}}_g||$ as sampling density indicator. Gradients are re-evaluated on this adaptive dataset including $\mathbf{X}_g$, $\mathbf{G_g}$, $\tilde{\mathbf{G}}_g$ using the same parameters.

The IsoUni approach of $\langle u'u' \rangle$ uses $r_c \approx 0.007$, resulting in \num{3915} bases. The adaptive collocation for both IsoAdapt and AnisoAdapt uses $r_c \in [0.002, 0.05]$, with $n_B = 783$ bases. For the AnisoAdapt approach, $\text{AR}_{\text{max}}=6$ is used. The collocation $\langle v'v' \rangle$ is similar and uses $r_c \approx 0.007$ for the IsoUni approach and  $r_c \in [0.003, 0.05]$ for the IsoAdapt and AnisoAdapt, with the same $\text{AR}_{\text{max}}=6$. All parameters are summarised in Table~\ref{tab:reg_param_case1}.

For each of the three approaches, different regressions are performed and categorised based on the source of the mean velocity field. We define $\langle u'_{\text{RBF}}u'_{\text{RBF}} \rangle$, $\langle v'_{\text{RBF}}v'_{\text{RBF}} \rangle$ as the fluctuations computed using the RBF regression of the mean velocities $\langle u \rangle$ and $\langle v \rangle$, for the corresponding approach. Likewise, we define $\langle u'_{\text{DNS}}u'_{\text{DNS}} \rangle$, $\langle v'_{\text{DNS}}v'_{\text{DNS}} \rangle$ as the fluctuations computed using the mean velocities from the DNS profile $\langle u \rangle _{\text{DNS}}$ and $\langle v \rangle _{\text{DNS}}$, linearly interpolated on the data points $\mathbf{X_*}$. The results for the streamwise and cross-stream fluctuation components are shown in Fig.~\ref{fig:case2_uu_results}.

The first noteworthy observation is the influence of the previous regression on the performance of the current one. In particular, for the uniform isotropic regression, where the estimation of $\langle u \rangle$ is strongly impacted by oscillations, a noticeable discrepancy appears between the $\langle u'u' \rangle$ regression obtained from the RBF data and that obtained from the DNS data. IsoAdapt and AnisoAdapt approaches estimate $\langle u \rangle$ sufficiently accurate that no such influence is visible in $\langle u'u' \rangle$. Likewise, the estimation of $\langle v \rangle$ is adequate across all approaches, leading to no observable differences between $\langle v'_{\text{RBF}}v'_{\text{RBF}} \rangle$ and $\langle v'_{\text{DNS}}v'_{\text{DNS}} \rangle$. 

Concerning the regression performance itself, hence focusing $\langle u'_{\text{DNS}}u'_{\text{DNS}} \rangle$ and $\langle v'_{\text{DNS}}v'_{\text{DNS}} \rangle$, it again becomes evident that the IsoUni approach fails to resolve the strong gradient. Oscillations appear for $y^+ > 100$ in both $\langle u'u' \rangle$ and $\langle v'v' \rangle$. The IsoAdapt approach mitigates the onset of oscillations by placing fewer—and therefore wider—bases in the region $y^+ > 100$. However, this does not improve the resolution of the gradient: a pronounced overshoot remains visible in $\langle u'u' \rangle$, and significant discrepancies with the DNS profile persist for $\langle v'v' \rangle$. In contrast, the AnisoAdapt approach provides markedly improved resolution thanks to the stretching of the bases. This leads to excellent agreement with the DNS solution for $\langle u'u' \rangle$ and only a slight deviation near $y^+ = 100$ for $\langle v'v' \rangle$, demonstrating that this method can indeed capture very sharp gradients without introducing oscillations. Another important aspect is the computational cost, which, in this case, is reduced by at least one third, from \num{1693} to \SI{1120}{\second} for $\langle u'u' \rangle$, and from \num{1756} to \SI{968}{\second} for $\langle v'v' \rangle$, when switching from the IsoUni approach to the AnisoAdapt approach, as detailed in Table~\ref{tab:computation_times}. It is worth reminding here that parameter optimisation is performed using an out-of-sample evaluation set, and the DNS is used for comparison purposes only.

\begin{figure}[t]
    \centering
    \includegraphics[width=1\linewidth]{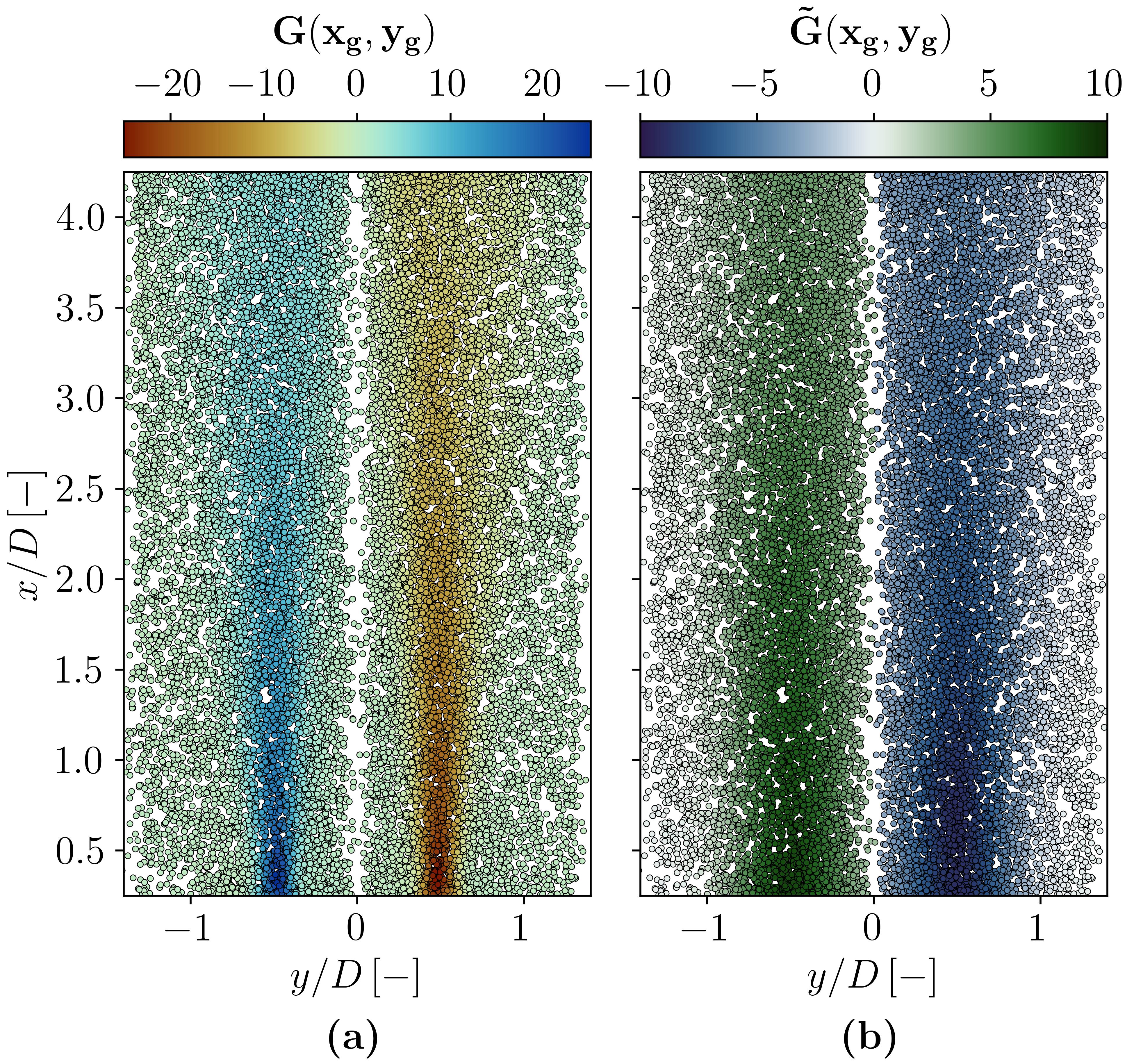}
    \caption{Test case 2: Estimated, raw gradient field $\mathbf{G}(\mathbf{x}_g, \mathbf{y}_g)$ (a) and smoothed gradient field $\tilde{\mathbf{G}}(\mathbf{x}_g, \mathbf{y}_g)$ with $\sigma_S=0.03$ (b)}
    \label{fig:case3_smoothed_grad}
\end{figure}

\subsection{3D velocimetry of a round jet}\label{sec4p3}

\begin{figure*}[t]
    \centering
    \includegraphics[width=1\linewidth]{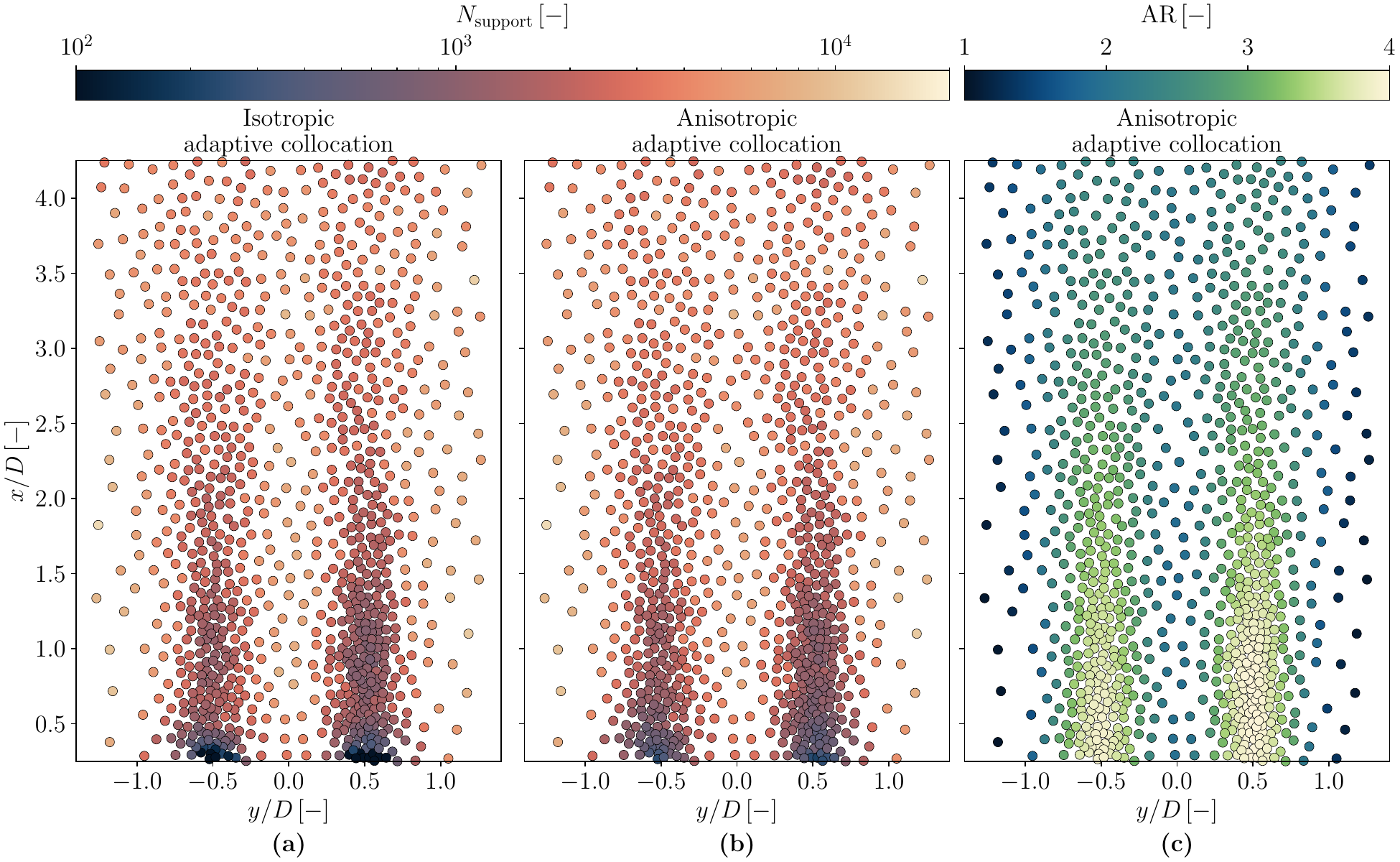}
    \caption{Test case 2: Spatial locations of the collocation points, coloured by the number of supporting points using the adaptive data sampling for the adaptive isotropic (a) and anisotropic adaptive basis (b). Aspect ratio (AR) of the anisotropic basis elements}
    \label{fig:case3_rbf_distri_AR}
\end{figure*}

The same methodology is employed for the second test case, using exclusively the adaptive data sampling. Starting with the regression of the axial velocity field, $\langle u \rangle$, the uniform collocation for the IsoUni approach is obtained with $r_c\approx0.009$, resulting in $n_B = \num{4820}$. Gradients are estimated over $\mathbf{X}_g$, of size $n_G = n_P/2 = \num{25000}$. The local regression is performed using $n_F = 1000$ neighbours. This initial gradient is depicted in Fig.~\ref{fig:case3_smoothed_grad}(a) together with the smoothed version (b) obtained with $\sigma_S=0.05$, which is then interpolated. From this interpolator, the collocation for the IsoAdapt and AnisoAdapt approaches are obtained with $r_c \in [0.008, 0.05]$ from the Poisson disk sampling algorithm. The resulting collocation points are presented in Figs.~\ref{fig:case3_rbf_distri_AR}(a) and (b) for the isotropic adaptive and anisotropic adaptive case, respectively. In both cases, the number of basis elements is fixed to $n_B=\num{964}$. As for the previous test case, the support of each basis remains similar, but particularly the bases in the shear layer at the inlet have larger support in the anisotropic case. This is because the particles move from bottom to top in the figure, meaning that at the bottom, less data points are available as the tracks are initialised from there. The anisotropic bases are stretched up to an aspect ratio of $\text{AR}_\mathrm{max} = 4$, as illustrated in Fig.~\ref{fig:case3_rbf_distri_AR}(c).

\begin{figure*}[t]
    \centering
    \includegraphics[width=1.0\linewidth]{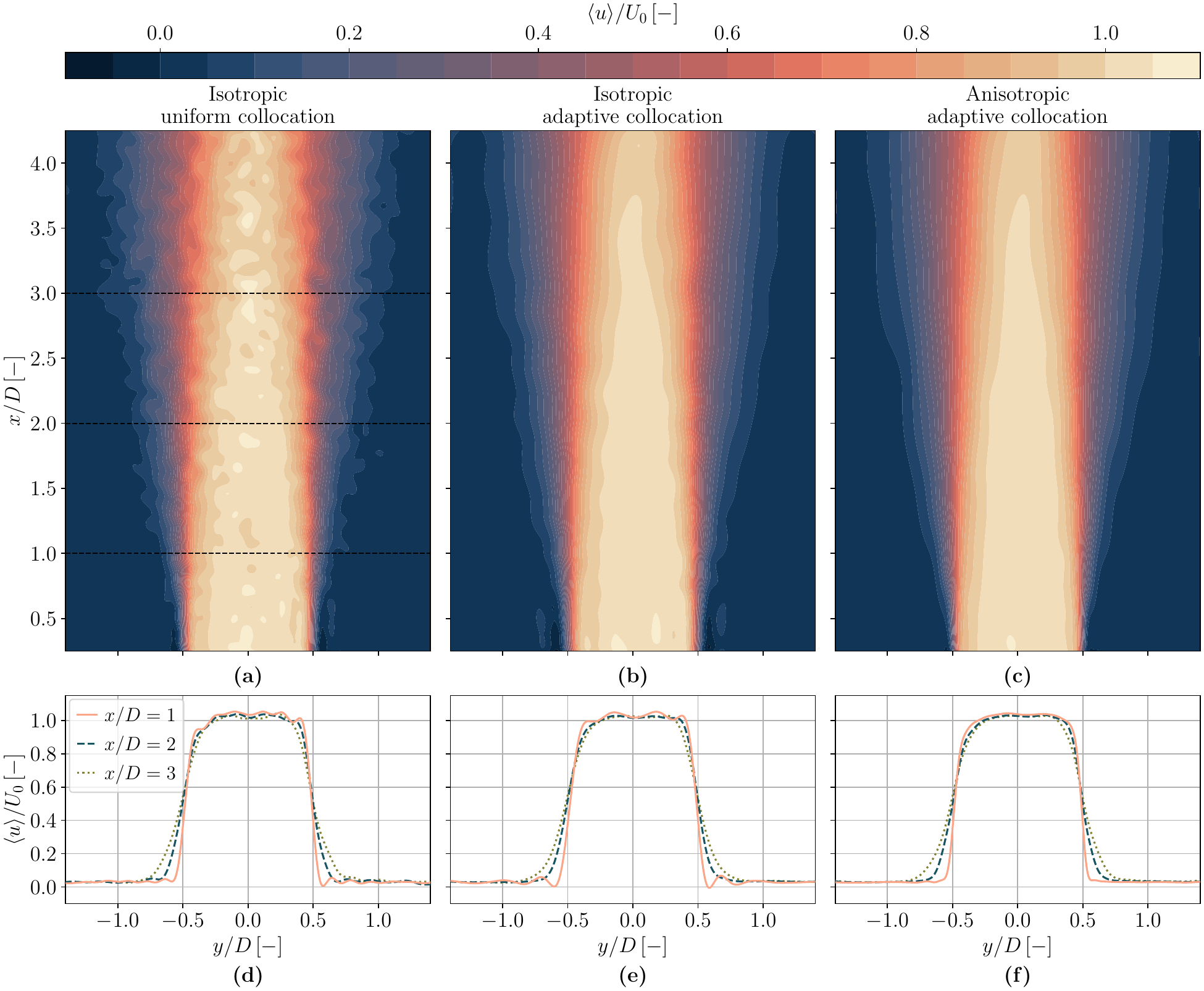}
    \caption{Test case 2: Streamwise mean velocity field $\langle u \rangle / U_0$ for the isotropic uniform (a), isotropic adaptive (b), and anisotropic adaptive (c) basis. The horizontal dashed lines in subfigure (a) correspond to three different slices of the velocity profile which are depicted in subfigures (d)-(f). The legend in subfigure (d) is shared among subfigures (d)-(f)}
    \label{fig:case3_u_results}
\end{figure*}

All approaches are solved using $\alpha_r=10^{-2}$ while no constraints are applied in this case. The optimised gradient penalty variables for the AnisoAdapt approach are summarized in Table~\ref{tab:reg_param_case2}. The optimisation of $\varepsilon_{g,x}$ and $\varepsilon_{g,y}$ here tends to remove, among the gradient points, those with a wrong estimation, negatively contributing to the general regression. \textcolor{black}{
The significant difference between the two thresholds reflects the anisotropic nature of the velocity gradients. Cross-stream gradients are significantly larger due to shear, whereas streamwise gradients are weaker and more sensitive to noise. 
Accordingly, the optimization selects thresholds that scale with the characteristic gradient magnitude in each direction.
}

For each of the three approaches, results are presented in the form of a 2D contour plot (Figs.~\ref{fig:case3_u_results}(a)-(c)). Moreover, a slice of the regression is taken at several downstream locations, corresponding to 1, 2, and 3 jet diameters from the nozzle exit (Figs.~\ref{fig:case3_u_results}(d)-(f)).

The IsoUni approach, as previously shown, struggles to resolve strong gradients. In this case, not only does an overshoot appear at $x/D = 1$, but the entire regression becomes contaminated by fluctuations introduced by the basis functions, whose scales are relatively small compared to the flow in this region. Once again, the size (and therefore the number) of these bases represents a trade-off between resolving the gradient and limiting these fluctuations. As shown in Fig.~ \ref{fig:case3_u_results}(b), the adaptive placement of the bases in the IsoAdapt approach eliminates these oscillations entirely, despite using only \SI{20}{\percent} of the original number of bases. However, the overshoot at one diameter downstream is not reduced; on the contrary, it increases slightly, as seen in Fig.~\ref{fig:case3_u_results}(e). Nevertheless, beyond two jet diameters downstream, the solution can be considered qualitatively reliable, with gradient levels that are adequately captured by the regression. The AnisoAdapt approach further improves performance, accurately reconstructing the velocity profile even at one jet diameter, where the gradient poses the greatest challenge. In addition, the corresponding regression field exhibits a noticeably smoother spatial distribution.

\begin{figure*}[t]
    \centering
    \includegraphics[width=1.0\textwidth]{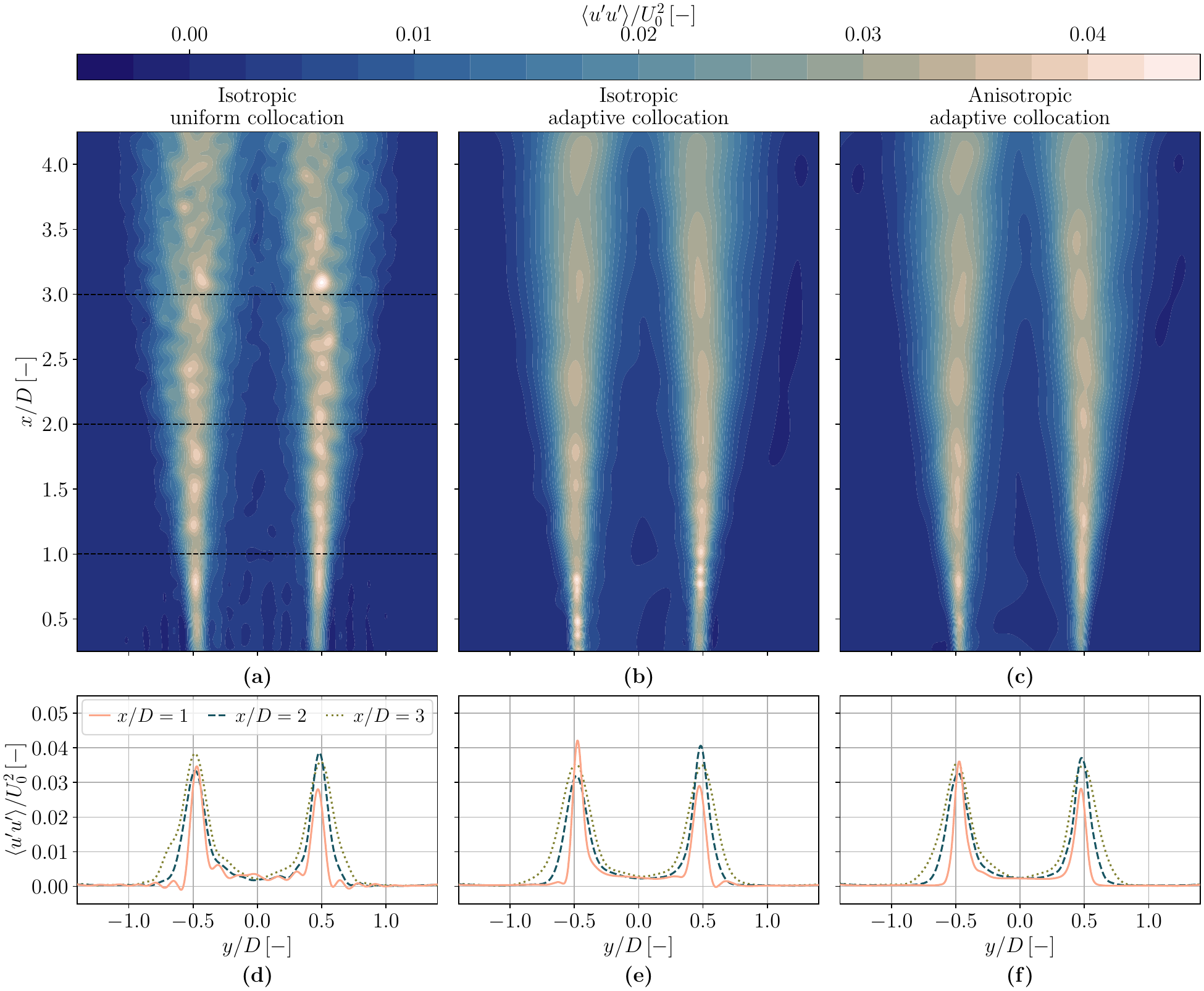}
    \caption{Test case 2: Streamwise Reynolds stress field $\langle u^\prime u^\prime \rangle / U_0^2$ for the isotropic uniform (a), isotropic adaptive (b), and anisotropic adaptive (c) basis. The horizontal dashed lines in subfigure (a) correspond to three different slices of the velocity profile which are depicted in subfigures (d)-(f). The legend in subfigure (d) is shared among subfigures (d)-(f)}
    \label{fig:case3_uu_results}
\end{figure*}

The stream-wise fluctuation component is computed according to \eqref{eq:u_prime} to obtain $\mathbf{u'}$. For IsoUni approach, a uniform collocation with $r_c \approx 0.007$ is used, resulting in a total of 4820 basis functions. The gradient estimation is performed on $n_G = n_P / 20 = \num{25000}$ points, with $n_F = 1000$. Smoothing is applied using $\sigma_S = 0.003$, and the resulting field is employed to determine the adaptive collocation, where $r_c$ varies within the range $[0.003,0.05]$. This procedure yields $n_B = 964$ basis functions. Finally, for AnisoAdapt approach, an anisotropic stretching with a maximum aspect ratio of $\text{AR}_{\max} = 4$ is applied. As with the stream-wise velocity, the regression results for the velocity fluctuations are presented for each approach, both as contour maps in Figs.~\ref{fig:case3_uu_results}(a)–(c) and as slices extracted at distances of 1, 2, and 3 jet diameters downstream. Consistent with the results obtained for the mean velocity $\langle u \rangle$, the IsoUni approach exhibits numerous fluctuations caused by the small basis functions. The IsoAdapt approach, despite using only \SI{20}{\percent} of the bases, successfully removes these artefacts, although the overshoots persist. In contrast, the AnisoAdapt approach again suppresses the oscillations entirely and yields a smooth transition to uniform grey levels without overshoot.

Examining the fluctuation intensity $\langle u' u' \rangle$, it is evident that the IsoUni and IsoAdapt approaches produce higher peak values. This behaviour stems from the poorer estimation of $\langle u \rangle$, which artificially inflates the fluctuation levels. Although the AnisoAdapt approach is capable of reconstructing sharper peaks than the other two methods, the resulting values remain lower, indicating a more accurate evaluation of $\langle u \rangle$. Overall, the AnisoAdapt approach consistently outperforms the traditional IsoUni formulations, while also requiring less computational time (\SI{738}{\second}) compared with the standard method (\SI{3026}{\second}).

\section{Conclusions and Perspectives}\label{sec.conclusions}

This work has presented an adaptive, anisotropic, and gradient-informed extension of the constrained radial basis function regression framework for the reconstruction of flow statistics from scattered data. By adapting the sampling density, the collocation layout, and the local shape of the basis functions to the variability of the flow, the method overcomes several limitations of classical isotropic RBF regression, notably its tendency to exhibit overshoots near sharp gradients and oscillations in smooth regions. The proposed formulation improves accuracy, robustness, and numerical stability while reducing computational cost, as the adaptive collocation naturally concentrates basis functions only where required.

The proposed methodology was evaluated using two demanding datasets: DNS of a turbulent channel flow and 3D-PTV measurements of an immersed jet. In both cases, the adaptive anisotropic strategy achieves high-resolution reconstruction of large velocity gradients, suppresses spurious oscillations, and remains robust under strong variations in sampling density. For the channel flow, the method yields substantially improved mean quantities compared with traditional approaches, which in turn leads to a more accurate estimation of turbulent fluctuations. The reconstructed statistics exhibit near-perfect agreement with the DNS reference, including in the near-wall region. Moreover, the combination of adaptive node placement and adaptive downsampling concentrates computational effort where it is most needed, resulting in an average reduction of approximately $50\%$ in computation time.

For the experimental jet, the methodology demonstrates strong robustness in a velocity field where gradient magnitudes vary significantly with both position and direction. The reconstruction provides qualitatively sharper and oscillation-free gradients throughout the domain. In contrast to the isotropic approach (IsoUni)—which exhibited pronounced oscillations in the velocity field at distances shorter than one jet-exit diameter—the adaptive anisotropic formulation successfully reconstructs the entire region without such artefacts.

Although the sampling strategy is already spatially guided by gradient magnitudes, ongoing work aims to incorporate turbulence intensity as an additional criterion for sampling density to further ensure statistical convergence. While turbulence is typically associated with strong gradients, small-scale gradients may be attenuated by the local regression procedure, preventing the adaptive downsampling from reflecting their presence. Integrating turbulence information will help address this limitation.

\section*{Acknowledgment}
D.R. is funded by the Fonds de la Recherche Scientifique (F.R.S.-FNRS) through the FRIA grant No. FC 58079. M.R. is funded by the Fonds de la Recherche Scientifique (F.R.S.-FNRS) through the FRIA grant No. FC57471. The experimental setup and M.A.M.'s contributions are supported by the European Research Council (ERC) under the Horizon Europe programme (Starting Grant No. 101165479, RE-TWIST). The data are visualized using perceptually uniform colormaps to prevent distortion of the data and exclusion of readers with color-vision deficiencies \citep{Crameri2020, Crameri2023}.





\bibliographystyle{spbasic}
\bibliography{bibliography.bib}

\appendix




\end{document}